\documentclass[12pt]{article}
\usepackage{amsfonts,amsmath}
\usepackage{amssymb}
\usepackage[hypertex]{hyperref} %LaTex
   \usepackage[utf8]{inputenc}
\makeatletter \@addtoreset{equation}{section} \makeatother

\def\theequation{\thesection.\arabic{equation}}

\newcommand{\tr}{\mathbb{D}}%   {\tr }
\newcommand{\Dt}{\rm d}%{\dr_{tot}}

\newcommand{\wmv}{\Omega}
\newcommand{\Go}{\Omega}
\newcommand{\bwmv}{{\bar{\Omega}}}
\def\Ex{{\E(\Omega)}}
\def\bEx{{\bar{\E}(\bar{\Omega})}}

\newcommand{\dga}{{\dot{\ga}}}
\newcommand{\f}{\frac }

\newcommand{\gt}{\tau}
\newcommand{\bgt}{\bar{\tau}}
 \newcommand{\eq}{\eqref}

\newcommand{\hhmt}{{{h}}}

\newcommand{\hmt}{{\vartriangle}}%{{\rm{d}^*}}%{{\mathfrak{h}}}% {{  {\Delta^{-1}}{}}}%
\newcommand{\be}{\begin{equation}}
\newcommand{\ee}{\end{equation}}
\newcommand{\bee}{\begin{eqnarray}}
\newcommand{\beee}{\begin{array}}
\newcommand{\eee}{\end{eqnarray}}
\newcommand{\eeee}{\end{array}}

\newcommand{\gn}{\nu}
\newcommand{\EE}{\mathcal{E}}
\newcommand{\EEE}{\mathbf{E}}
\newcommand{\bEE}{\bar{ \mathcal E }}

\newcommand{\gx}{\xi}
\newcommand{\gr}{\rho}
\newcommand{\ga}{\alpha}
\newcommand{\pa}{{\dot{\ga}}}
\newcommand{\pb}{{\dot{\gb}}}

\newcommand{\pga}{{\bar{\gamma}}}
\newcommand{\pn}{{\dot{\nu}}}

\newcommand{\gb}{\beta}
\newcommand{\gga}{\gamma}

\newcommand{\M}{{\cal M}}

\newcommand{\E}{{\cal E}}

\newcommand{\rhs}{{\it r.h.s.} }
\newcommand{\rhss}{{\it r.h.s.'s} }

\newcommand{\ie}{{\it i.e.,} }
\newcommand{\ls}{\!\!\!\!\!\!}

\newcommand{\gd}{\delta}

\newcommand{\gep}{\epsilon}

\newcommand{\gs}{\sigma}

\newcommand{\bz}{\bar z}
\newcommand{\go}{\omega}

\newcommand{\by}{{\bar{y}}}
\newcommand{\brr}{{\bar{r}}}

\newcommand{\bk}{{\bar{k}}}

\newcommand{\bp}{{\bar{p}}}

\newcommand{\q}{\,,\qquad}

\newcommand{\dgb}{{\dot{\beta}}}

\newcommand{\ba}{{\overline{a}}}

\newcommand{\nn}{{\nonumber}}

\newcommand{\half}{\frac{1}{2}}

\newcommand{\p}{\partial}

\newcommand{\D}{{\cal D}}

\newcommand{\ff}{\frac}

\newcommand{\dr}{{\rm d}}

\begin{document}

\begin{flushright}
 {\small FIAN/TD/14-23}
\end{flushright}
\vspace{1.7 cm}
\begin{center}
 \end{center}

\begin{center}
{\large\bf Moderately non-local  $\eta \bar\eta$ vertices in the
$AdS_4$ higher-spin gauge theory}

\vspace{1 cm}

{\bf  O.A.~Gelfond}\\
\vspace{0.5 cm}
{\it
 I.E. Tamm Department of Theoretical Physics, Lebedev Physical Institute,\\
Leninsky prospect 53, 119991, Moscow, Russia}

 \end{center}

\vspace{1.2cm}

\vspace{0.4 cm}

\begin{abstract}

    A new concept
of  moderate non-locality in higher-spin gauge theory is introduced.
 Based on the recently proposed differential  homotopy approach, a moderately non-local scheme, that is
 softer than those resulting from the shifted homotopy approach
 available in the literature so far,   is worked out in the mixed $\eta\bar\eta$ sector
 of the  Vasiliev higher-spin theory.
   To calculate moderately non-local
vertices  $\Upsilon^{\eta\bar \eta}(\go, C,C,C)$
for all ordering of the fields  $\go$ and $C$
we apply an interpolating homotopy, that respects the moderate non-locality in the
 perturbative analysis of the  higher-spin equations.

\end{abstract}

\newpage
\tableofcontents

\newpage

\section{Introduction}
\label{intro}

\subsection{Preliminaries}\label{prel}

Higher-spin (HS) gauge theory describes interacting systems of massless fields of all spins,
 resulting from a nonlinear deformation of the Fronsdal theory of free  HS fields
\cite{Fronsdal:1978rb}.
Such  theories  play a role in various contexts from holography \cite{Klebanov:2002ja} to cosmology
\cite{Barv}. A useful way of description of HS dynamics in $AdS_4$ is provided by the
generating Vasiliev system of HS equations  \cite {Vasiliev:1992av}. The latter
contains a free complex parameter
 $\eta$. Reconstructing on-shell HS vertices order by order one obtains vertices
proportional to various powers of $\eta$ and $\bar{\eta}$.

Since the  HS gauge theory contains
infinite tower of gauge fields of all spins and the number of space-time derivatives increases with the spins
of fields  in the vertex \cite{Bengtsson:1983pd}-\cite{Fradkin:1991iy},
 the theory
exhibits certain degree of non-locality. The level of non-locality of HS gauge theory is
debatable in the literature.

In the papers
\cite{Vasiliev:2017cae}-\cite{Gelfond:2021two}    vertices in the holomorphic  (anti-holomorphic) sector   up to  $\eta^2$   ($\bar{\eta}^2)$,
 were reconstructed from the generating Vasiliev system  in the spin-local form.
 (See also \cite{Didenko:2022qga} for a higher-order extension of these results.)
The shifted homotopy approach used in \cite{Vasiliev:2017cae}-\cite{Didenko:2020bxd} demands careful choice of the
homotopy scheme  compatible with the spin-locality of the vertices.

 Being  efficient in the
lowest orders, the original shifted homotopy approach  turns out to be less powerful at higher orders.
This way,
it has not been yet possible to find spin-local
  vertices  in the so called mixed $\eta  \bar{\eta} $  sector of equations for zero-form fields.

From the perspective of HS theory in the bulk it is hard to identify the minimal level
of non-locality of the theory unless a constructive scheme that supports
some its specific  level is presented.  The aim of this paper is to present such a scheme
that supports a moderate non-locality
of the HS theory in the mixed $\eta\bar\eta$
sector, that is less non-local than those resulting from the shifted homotopy approach
 available in the literature so far. Specifically, we will use the
differential homotopy approach proposed recently by Vasiliev in \cite{Vasiliev:2023yzx}
 to obtain   moderately non-local vertices
$\Upsilon^{\eta\bar \eta}(\go,C,C,C)$   for the zero-form
equations in the mixed sector.

Since the moderately non-local vertices obtained in this paper minimize the level
of non-locality of the known HS vertices, it would be interesting to compare it
with the level of non-locality of the vertices obtained in \cite{Sleight:2017pcz} via holographic duality based on the Klebanov-Polyakov conjecture
\cite{Klebanov:2002ja} (see also \cite{Bekaert:2015tva}-\cite{Neiman:2023orj}). {\it A priori}, there are two options:

 $(i)$ Moderately non-local vertices may have the same  (or even worse) level of nonlocality than that deduced in \cite{Sleight:2017pcz}.

 $(ii)$ Moderately non-local vertices of this paper
 may be less nonlocal than  those of \cite{Sleight:2017pcz}.

 The option $(i)$ is in fact inconclusive since it is not guaranteed
that there is no  better scheme allowing to soften further
the level of vertex non-locality. On the other hand, the option $(ii)$
would imply that the HS holographic duality has to be modified one way or
another, for instance along the lines of \cite{Vasiliev:2012vf}. Though the
analysis of this issue is very interesting, it is not
straightforward because of the  difference between the formalisms
underlying the space-time analysis of  \cite{Sleight:2017pcz} and
the unfolded analysis of this paper in terms of auxiliary spinor
variables.  Hence it is postponed for  the future study.

The paper is organized as follows.
In Section \ref{HSsketch} we recall the form of HS  equations.
In Section \ref{difhom}  the Vasiliev concept of differential homotopy
and  the  Ansatz for the linear in $\eta$   deformations of \cite{Vasiliev:2023yzx} are recalled.
In Section \ref{Ansatz} the  Ansatz for the linear in $\eta\bar \eta$   deformations
is  introduced,
as a straightforward generalization of that of \cite{Vasiliev:2023yzx}.
In Section \ref{LOCALITY} we briefly discuss  a locality issue and introduce a notion of
'moderate spin-non-locality' (MNL), also introducing 'interpolating homotopy' (IH)  that respects MNL. In Section \ref{MNLB3} the  derivation of the   MNL  $B_3$ is
considered   in detail.
 In Section \ref{MNLvert} the resulting MNL vertices $\Upsilon^{\eta\bar\eta}(\go, C,C,C)$  are
 introduced.
Conclusions   are summarized in Section \ref{Con}.
  Appendices
  {\bf A} - {\bf C}  collect previously known
  results of the lowest-order computations while Appendices
  {\bf D} and  {\bf E}  contain  vertex   $\Upsilon_{ \go CC  C} $ and $\Upsilon_{ C \go CC  } $
  calculation details,  respectively.

\section{Higher-spin equations}
\label{HSsketch}

\subsection{Original form}

The nonlinear HS equations of \cite{Vasiliev:1992av}  \begin{align}
&\dr_x W+W*W=0\,,\label{HS1}\\
&\dr_x S+W*S+S*W=0\,,\label{HS2}\\
&\dr_x B+[W,B]_*=0\,,\label{HS3}\\
&S*S=i(\theta^{A} \theta_{A}+ B*(\eta \gga +\bar \eta \pga ))\,,
\label{HS4}\\
&[S,B]_*=0\,\label{HS5}
\end{align}
 reproduce  field equations on dynamical HS fields in any gauge and choice of field variables.
 The field $B(Z;Y;K|x)$ is a zero-form, $x$ are space-time coordinates,
    $Z_A=(z_{\ga},
\bar z_{\pa})$, $Y_A=(y_{\ga}, \bar y_{\pa})$  are
auxiliary commuting spinor variables ($\ga,\gb=1,2$; $\pa,\dgb =1,2$),
$\eta $ is a free complex parameter ($\bar \eta$ is its complex conjugate) and
$K=(k,\bar k)$ are involutive Klein operators obeying
\be\label{hcom}
\{k,y_{\ga}\}=\{k,z_{\ga}\}=0\,,\qquad [k,\bar y_{\pa}]=[k,\bar
z_{\pa}]=0\,,\qquad k^2=1\q [k\,,\bar k]=0\,.
\ee
Analogously for $\bar k$.

  The field  $W(Z;Y;K|x)$
   is a space-time  one-form, \ie $W= \dr x^\nu W_\nu$,
while  $S(Z;Y;K|x)$   is a one-form in $Z$ spinor directions, \ie $S= \theta^\ga S_\ga+ \bar{\theta}^{\pa}  S_\pa$,  $ \theta^\ga:= \dr z^\ga,
 \bar{\theta}^{\pa}:= \dr \bz^\pa$. The wedge symbol is implicit in this paper since all
 products are exterior.

The star product is
\be\label{star}
(f*g)(Z, Y)=\ff{1}{(2\pi)^4}\int dU dV f(Z+U; Y+U)g(Z-V;
Y+V)\exp(iU_{A}V^{A})\,.
\ee
Indices are raised and lowered by  the symplectic form $\gep_{AB}=(\gep_{\gb\ga}, \gep_{\pb\pa}) $,
\be
 X^{A}=\gep^{AB}X_{B}\q X_A=X^{B}\gep_{BA}\,.
 \ee

 Elements $\gga$ and $\pga$,
  \be\label{klein}
\gga=\exp({iz_{\ga}y^{\ga}})k\theta^{\ga} \theta_{\ga}\,,\qquad
\pga =\exp({i\bar{z}_{\pa}\bar{y}^{\pa}})\bar
k\bar\theta^{\pa}\bar\theta_{\pa}\,
,\ee
 are central with respect to the star product
 since $\theta^3=\bar\theta^3=0$.

Following \cite{Vasiliev:1992av}, to analyse equations (\ref{HS1})-(\ref{HS5}) perturbatively  one
starts with the vacuum solution
\be
B_0=0\label{B0}\q
S_0=\theta^A Z_{A}=\theta^\ga z_{\ga}+\bar{\theta}^{\pa}\bar
z_{\pa}\,.
 \ee
Plugging this into \eqref{HS1}-\eqref{HS5} and using  that
\be
\label{S0}
[S_0\,,]_* = -2i \mathrm{d}_Z\q \mathrm{d}_Z:= \theta^A \frac{\p}{\p Z^A}\,,
\ee
one finds that $W_0$ should
be $Z$-independent, $W_0=\go(Y; K|x)$, and satisfy \eqref{HS1}.
Similarly, at the next order one gets $B_1=C(Y; K|x)$ from $[S_0,
B_1]=0$ and that $C$ satisfies \eqref{HS3}. This way one reconstructs the first
terms on the \rhss of the unfolded equations of the form originally proposed in
\cite{Vasiliev:1988sa}
\begin{equation}\label{HSsketch1}
\ls\dr_x \omega+\go\ast \go=\Upsilon^{\eta }(\go,\go,C)+\Upsilon^{ \bar \eta}(\go,\go,C)
+\Upsilon^{\eta \eta}(\go,\go,C,C)+\Upsilon^{\bar\eta\bar \eta}(\go,\go,C,C)
+\Upsilon^{\eta\bar \eta}(\go,\go,C,C)\ldots,
\end{equation}
\begin{equation}\label{HSsketch2}\ls\dr_x C+[\omega\,, C]_ \ast=
\Upsilon^\eta(\go,C,C)+\Upsilon^{\bar \eta}(\go,C,C)+
\Upsilon^{\eta  \eta}(\go,C,C,C)
+\Upsilon^{\bar\eta\bar \eta}(\go,C,C,C)+\Upsilon^{\eta\bar \eta}(\go,C,C,C)\ldots.
\end{equation}
 As in \cite{Vasiliev:1988sa},
the resulting perturbative
expansion is in powers of the zero-forms $C$.

To obtain dynamical equations \eqref{HSsketch1}, \eqref{HSsketch2} one should plug obtained
$B_i, W_i$ into Eqs.~\eqref{HS1}, \eqref{HS3}. For instance,
 \eqref{HS3} up to the third order in $C$-field is
  \begin{equation}\label{dxC=}
\dr_x C+[\go,C]_\ast=-\dr_x B_2-[W_1,C]_\ast-\D  B_3 -[W_1,B_2]_\ast-[W_2,C]_\ast \q
\end{equation}
where
\be
\label{cad}
\D A: =\dr_x A +[\go\,, A]_*\,.
\ee
For more detail we refer the reader to the review \cite{Vasiliev:1999ba}.

  \subsection{Free equations in $AdS_4$}

$AdS_4$ vacuum one-form connection $W_0$  is
\be
\label{AdS}
 W_0 = \half w^{AB} (x) Y_A Y_{B}\q
\dr w^{AB} +w^{AC} C_{CD} w^{DB}=0\q
\ee
where  $C_{AB}$  ($A,B\ldots =1\,,\ldots4$) is the $sp(4)$ invariant form,   $w^{AB}=(\go^{\ga\gb},\bar \go^{\dga\dgb}, e^{\ga\dga})$
describes Lorentz connection, $\go^{\ga\gb},\bar \go^{\dga\dgb}$, and vierbein, $e^{\ga\dga}$, with two-component
spinor indices $\ga=1,2$, $\dga=1,2$ (at the convention $A^\ga =\epsilon^{\ga\gb}A_\gb$,
$A_\ga =A^\gb\epsilon_{\gb\ga}$).
The  unfolded system for  free massless   fields  $\go(y,\by; K | x)$ and   $C(y,\by;K | x)$
  reads as \cite{Vasiliev:1988sa}
\bee \label{tDO} && R_1(y,\overline{y};K\mid x) = \frac{i}{4}\left ( \eta \overline{H}^{\dga\pb}
\bar \p_\dga \bar \p_\dgb
{C}(0,\overline{y};K\mid x)k + \bar \eta H^{\ga\gb}\p_{\ga}\p_{\gb}
{C}(y,0;K\mid x)\bar k\right ) , \\ \label{tDC}
 &&
\tilde{D}_0 C(y,\overline{y};K\mid x) =0\,, \label{DC}
\eee
where
\be
\p_\ga :=\f{\p}{\p y^\ga}\q \bar\p_\dga :=\f{\p}{\p \by^\dga}\,,
\ee
\be H_{\ga\gb}:= e_\ga{}_\dga e_\gb{}^\dga\q
\overline H_{\dga\dgb}:= e_\ga{}_\dga e^\ga{}_\dgb\,,
\ee
\be
\label{R1}
R_1 (y,\bar{y};K\mid x) :=D^{ad}_0\omega (y,\bar{y};K\mid x)\qquad
D^{ad}_0  = D^L  -
 e^{\ga\pb}\Big(y_\ga \bar \partial_\pb
+ {\p_\ga}\bar{y}_\pb\Big)\,,
\ee
$$
D^L   = \dr_x  -
\Big(\go^{\ga\gb}y_\ga {\p_\gb} +
\bar{\go}^{\dga\pb}\bar{y}_\dga\bar \p_\dgb \Big)\,,
$$
\be
\label{TD}
\tilde{D}_0  = D^L  + e^{\ga\pb}
\Big(y_\ga \bar{y}_\pb +\p_\ga
\bar \p_\pb\Big)\,.
\ee
The massless fields obey
\be
\go(y,\by;-k,-\bar k\mid x) = \go(y,\by;k,\bar k\mid x)\q
C(y,\by;-k,-\bar k\mid x) = - C(y,\by;k,\bar k\mid x)\,.
\ee

 System (\ref{tDO}), (\ref{tDC}) decomposes
into  subsystems of different spins, with a  spin $s$  described by
the one-forms $ \omega (y,\bar{y};K| x)$ and zero-forms $C (y,\bar{y};K| x)$ obeying
\be
\label{mu}
\omega (\mu y,\mu \bar{y};K\mid x) = \mu^{2(s-1)} \omega (y,\bar{y};K\mid x)\q
C (\mu y,\mu^{-1}\bar{y};K\mid x) = \mu^{\pm 2 s}C (y,\bar{y};K\mid x)\,,
\ee
where  $+$ and $-$   correspond to helicity $h=\pm s$ selfdual and anti-selfdual parts
of the generalized Weyl tensors $C (y,\bar{y};K| x)$.

We consider equations (\ref{tDC}) on the gauge invariant zero-forms $C$
$$
C(Y;K|x)=\sum^1_{A=0}\sum_{n,m=0}^\infty\f{1}{2 n!m!}
C^{A\,1-A}_{\ga_1\ldots \ga_n\,,\dga_1\ldots \dga_m}(x)
y^{\ga_1}\ldots y^{\ga_n}
\bar{y}^{\dga_1}\ldots \bar{y}^{\dga_m}k^A \bar k^{1-A}\,.
$$
Spin-$s$  zero-forms are $C^{A\,1-A}_{\ga_1\ldots \ga_n\,,\dga_1\ldots \dga_m}(x)$ with
$n-m=\pm 2s\,.
$
Eq.~(\ref{DC}) rewritten in the form
\be\label{xyy}
D^L C  = e^{\ga\dgb}{ \frac{\partial^2}{\partial y^\ga
\partial \bar{y}^\pb}} C  +
\mbox{ lower-derivative and nonlinear terms}\,
\ee
(discarding   indices $A$)
{implies that higher-order terms in} $y$ {and} $\bar y$ in the zero-forms $C(y,\by| x)$
 describe higher-derivative descendants of the primary components $C(y,0| x)$
and  $C(0,\by| x)$ relating second derivatives in $y,\bar y$ to
the $x$ derivatives of  $C (Y;K|x)$ of lower degrees
in $Y$.   Generally, $C_{\ga_1\ldots \ga_n\,,\dga_1\ldots \dga_m}(x)$ contains
$\f{n+m}{2}-\{s\}$ space-time  derivatives of the  spin-$s$ dynamical
field.   As a result, the zero-forms
$C$ in the HS vertices may induce infinite towers of derivatives and, hence,
non-locality.

\section{Vasiliev's differential homotopy approach}
\label{difhom}
Here we recall the concept of differential homotopy of \cite{Vasiliev:2023yzx} and  the Ansatz for
(anti)holomorphic deformation linear in ($\bar \eta$)$\eta$   and discuss its $\eta\bar\eta$
generalization  used in this paper.

\subsection{Differential homotopy }
\label{Differential homotopy}

Shifted contracting homotopy $\hmt_{q}$  and cohomology projector $\hhmt_{q }$ act as follows
\cite{Gelfond:2018vmi}
\be\label{homint0} \hmt_{ q} \phi(Z,Y, \theta) =\int_0^1 \f{dt }{t } (Z+ q)^A\f{\p}{\p \theta^A}
 \phi( t  Z-(1-t ) q,t \,\theta)\q\hhmt_{q} \phi(Z,Y, \theta)= \phi(-q,Y,0)\,
\ee
obeying the standard resolution of identity
\be\label{newunitres}
\left\{ \dr_Z\,,\hmt_{q }\right\} +\hhmt_{q }=Id\, .
\ee

Here a shift $q$ must be independent of $Z$ and $t$ but
can depend on some parameters and/or integration variables.
Moreover, further   contracting homotopies lead to multiple integrals
$\int  {d t^1 }\int {d t^2 } \ldots$.
All of these
  parameters    were    interpreted in  \cite{Vasiliev:2023yzx}
 as additional coordinates  $t^a$
 with the total differential
 \be
\label{td}
\dr= \dr_Z +\dr_t\,,
\ee
\be
\dr_Z = \theta^A \ff{\p}{\p Z^A}\q \dr_t = dt^a\ff{\p}{\p t^a}\,,
\ee
where $\theta^A$ and $dt^a$ are anticommuting differentials and the homotopy coordinates $t^a$  belong to a unite hypercube,
\be
0\leq t^a\leq 1\,.
\ee
In these terms,  perturbative
equations to be solved acquire  the form
\be
\label{totg}
\dr f(Z,t ,\theta,dt ) = g(Z,t ,\theta,dt )\q \dr g(Z,t ,\theta,dt )=0\,,\ee
where the second one expresses the compatibility of the first with $\dr\dr=0$.

Functions like $f$ and $g$ contain theta and delta
functions like
$\theta(t^a) \theta(1-t^a)$ restricting the $t$ integration to a locus inside
a unit hypercube.
Physical fields and equations in HS theory
are  supported by $\dr_t$ cohomology
carried by the integrals over $t^a$.

  Differential homotopy   is based on the removal of the integrals.
Namely, following \cite{Vasiliev:2023yzx} let \be\label{dfint}
\dr_Z f_{int} = g_{int}\q f_{int} = \int_{\M} f(Z,\theta, t, d t)\q
g_{int} = \int_{\M} g(Z,\theta, t,d t)\,,
\ee
resulting in
\be
\dr_Z f = g +\dr_t  h +g^{weak}\,,
\ee
where $g^{weak}$ does not contribute to the integral because its degree $\deg g^{weak}$ as
a form in $\M$ differs
from $\dim \M$. Setting $g^{weak} = \dr_Z h - \dr_t f $ (taking into account that
$\deg h=\dim \M-1$ and $\deg f =\dim \M$) and replacing $f\to f-h$, we obtain (\ref{totg}).

  $\M$ can be treated as  $\mathbb{R}^n$.
  Following \cite{Vasiliev:2023yzx}, for $\int_{\M}$ we  use
 notation
$  \int_{t ^1}\ldots \int_{t^k}  := \int_{ t ^1\ldots t ^k} \, $ with the convention that it
 is totally antisymmetric
in  $t^a$.
 Though the integrals are removed from the equations,
 to avoid a sign ambiguity due to (anti)commutativity of
differential forms,   every differential expression
is   accompanied with  integrals   $\int_{t ^1\ldots t ^k}$ that   can
be written anywhere in the expression for the differential form to be integrated
with the convention
\be \label{MVintconv}
\dr \int_{t ^1\ldots t ^k} = (-1)^k \int_{t^1\ldots t^k} \dr  \,.
\ee

 \subsection{Differential homotopy Ansatz for the  $\eta$   deformation}
\label{AnsatzMV}
As shown in \cite{Vasiliev:2023yzx}  direct computation within the differential homotopy approach gives the following
 form for the  lowest order holomorphic deformation linear in $\eta$
 in the perturbative analysis

\be
\label{feta}
f_\mu=  \eta\int_{u^2 v^2 \tau\gs\gb\rho }\ls\mu(\tau,\gs,\gb,\rho,u,v ) \dr \Go^2
\Ex G_l(g(r)) \big |_{r=0}\,,
\ee
 \be\label{Omega2}
\dr \Go^2 := \dr \Go^\ga \dr \Go_\ga\,,
\ee
\be\label{Exp}
\Ex:=   \exp i \big ( \Go_\gb(y^\gb +p^\gb_++u^\gb)+u_\ga v^\ga -\!\sum_{k\geq j>i\geq 1} p_{i\gb}
p_j^\gb \big)
\ee
\be
G_l(g):= g_1(r_1)\ldots g_l(r_l) k \,,
\ee
\be
p_{+\ga} = \sum_{i=1}^k p_{i\ga}\q p_{j\ga} = -i \frac{\p}{\p r^j{}^\ga}\,,
\ee
$g_i(y)$ are some functions of $y_\ga$ (e.g., $C(y)$ or $\go(y)$)
(anti-holomorphic variables $\bar y_\dga$, Klein operators
$K=(k,\bk)$ and
the antiholomorphic star product $\bar *$ are implicit).
\be\label{fulldif}
\dr =d\tau\f{\p}{\p \tau}
 +d\rho\f{\p}{\p \rho}+ d\gs_i \f{\p}{\p \gs_i} +
d\gb\f{\p}{\p \gb} +du^\ga\f{\p}{\p u^\ga} + dv^\ga\f{\p}{\p v^\ga}\,
\ee
and
\be
\label{muv}
\mu(\tau,\gs,\gb,\rho,u,v ) =\mu(\tau,\gs,\gb,\rho)d^2 u d^2 v
\q\ee
 where $du^\ga$ and $dv^\ga$ are anticommuting differentials  ,
\be
\label{meas}
d^2 u =du^\ga du_{\ga}\q d^2 v =dv^\ga dv_{\ga}
\q\int d^2 u d^2 v \exp i u_\ga v^\ga =1\,,
\ee
$ \Go_\ga  $ has the form
\bee\label{Omega=}
\label{sW}
 \Go_\ga  := \tau z_\ga - (1-\tau) (p_\ga (\gs) -\gb v_\ga +\rho(  y_\ga +p_{+\ga} +u_\ga )) \,,
\eee
where
\be
p_\ga(\gs)=\sum_{i=1}^k p_{i\ga} \gs_i
\ee
with some  parameters $\gs_i$. We use the convention of  \cite{Vasiliev:2023yzx} that it does not matter where the symbol of integral
is situated;  the integration over $d^2 u$ and $d^2 v$ in (\ref{feta}) also accounts for
 the $u,v$--dependent measure factor $\dr\Go^2$.

The   measure $\mu \dr\Go^2$ may contain so called { \it weak}
 terms   that do not contribute under the integration if the number of integrations
 does not match the number of respective differentials.
This   issue plays important role in the computations of \cite{Vasiliev:2023yzx}.

Due to the identity $(\dr\Go)^3=0$ being a consequence of the anticommutativity
of $\dr \Omega_\ga$ and two-componentness of the spinor indices $\ga$,
formula (\ref{feta}) has the following remarkable property  \cite{Vasiliev:2023yzx}
\be\label{dGOGO=0}\!\!
\dr[ \dr^2  u\dr^2 v \dr\Go^2\Ex]=\dr \Big (\dr^2  u \dr^2 v \dr \Go^2
\exp i \big ( \Go_\gb(y  \!+\!p_+\!+\!u)^\gb\!+\!u_\ga v^\ga\! -\!\!\!\!\sum_{k\geq j>i\geq 1}
\!\!\!p_{i\gb}
p_j^\gb \big)\Big ) =0\,.
\ee
As a result,  \be
\label{dmu}
\dr f_\mu = (-1)^N f_{\dr \mu}\,,
\ee
where $N$ is the number of the integration parameters $\tau, \gs_i, \gb,\rho$.
 By virtue of (\ref{dmu}),  equation (\ref{totg}) amounts to
\be
f_{\dr \mu_f} = g_{\mu_g}\,.
\ee
 This demands
\be
\dr \mu_f \cong \mu_g\,,
\ee
where  $\cong$ denotes the weak equality up to possible weak terms, that do not contribute
under the integrals in $f_{\dr \mu_f}$ and $ g_{\mu_g}$.
Since $g$ in (\ref{totg}) is
$\dr$ closed
\be
\dr \mu_g\cong 0\,.
\ee
In most cases this implies that
\be
\mu_g \cong \dr h_g
\ee
allowing to set
\be
\mu_f = h\,.
\ee

\subsection{Ansatz for the  $\eta\bar \eta$   deformations}
\label{Ansatz}   In this paper we use a particular case of
 Vasiliev's Ansatz (\ref{feta})
with $\gr=\gb=0$   allowing to discard  the dependence on $u$ and $v$,
that trivializes at $\gb=0$.

Firstly, recall that HS equations  remain consistent
with the fields $W$ and $B$ valued in any associative algebra \cite{Vasiliev:1988sa}.
As a result, the components of $W$ and $B$ do not commute and different orderings of the fields
can be considered independently.
Hence, functions $G_l(g,K) $ under consideration with $l=3$ and $l=4$, being at least linear in $\go$, can be  represented as a sum of expressions with
  different positions of $\go$.
  For the future convenience we denote arguments of
$\go$ as $r_0,\brr_0$ for any ordering.
 Namely, for $l=3,4$
  \bee \label{G34}
 G_l(g)=\left\{
\beee{l  }
C(r^1,\brr^1) C(r^2,\brr^2)  C(r^3,\brr^3)k\bar k \,,\\\\
\go(r^0,\brr^0)C(r^1,\brr^1) C(r^2,\brr^2)  C(r^3,\brr^3)k\bar k\,,\\ \\
C(r^1,\brr^1)\go(r^0,\brr^0) C(r^2,\brr^2)  C(r^3,\brr^3)k\bar k\,, \\\\
C(r^1,\brr^1) C(r^2,\brr^2) \go(r^0,\brr^0) C(r^3,\brr^3)k\bar k \,,\\\\
C(r^1,\brr^1) C(r^2,\brr^2)  C(r^3,\brr^3) \go(r^0,\brr^0)k\bar k\,.\\
  \eeee\right.\,
\eee
To simplify formulae we will use shorthand notations $\go C  CC$ instead  of
\\$\go(r^0,\brr^0)C(r^1,\brr^1) C(r^2,\brr^2)  C(r^3,\brr^3)|_{r^i=\brr^i=0}$  {\it etc}.

In this paper, we introduce Ansatz   in the bilinear $ \eta\bar\eta$ deformation
with
 \bee
\label{fetabeta}&&
F=\sum_i F^i \quad\mbox{ where }\quad
F^i=   \eta\bar\eta \int\limits_{\gt \bgt \gs(n) } \mu^i( \gt,\bgt, \gs )
   \EEE(\wmv^i |\bwmv^i\,)    G_l(g )        \eee
   with the  some compact  measure factors $ \mu^i( \gt,\bgt, \gs) $,  $G_l(g)$  \eq{G34},
\be \label{EEE}
\EEE(\wmv^i,\bwmv^i)=(\dr\wmv^i)^2
(\dr   \bwmv^i)^2\E(\Go^i)  \bar\E(\bar\Go^i)
\ee
with
\be\label{ExpO}
\Ex:=   \exp i \big( \Go_\gb(y^\gb +p^\gb_+ )  -\!\sum_{3\geq j>i\geq 1} p_{i\gb}
p_j^\gb -\!\sum_{3\geq j \geq 1} s_j p_{0\gb} p_j^\gb
\big)\,,\ee
\be\label{bExp}
 \bEx :=   \exp i \big ( \bar{\Go}_\pb(\by +\bp _+  )^\pb  -\!\sum_{3\geq j>i\geq 1}
 \bp_{i\pb}
\bp_j^\pb -\!\sum_{3\geq j \geq 1} \bar s_j \bp_{0\pb} \bp_j^\pb\big)\,,\ee
  \be\label{ Omegai}
 \Go^i_\ga  := \gt  z_\ga - (1-\gt)  a^i{}^j(\gs) \,p_j{}_\ga
\q
 \bar\Go^i_\pa  := \bgt \bz_\pa - (1-\bgt)   \ba{\,}^i{}^j(\gs) \, \bp_j{}_\pa \q
\ee  where  $s_j,\,\bar s_j$ are sign factors that
depend on the ordering of fields $C$ and $\go$   \eq{G34},
 $\gs $ are integration parameters and  $ a^{ij}(\gs) $, $\bar a^{ij}(\gs) $ are
some rational  functions that satisfy inequalities $ |a^{ij}(\gs)|\le1 $, $ |\bar a^{ij}(\gs)|\le1 $. The notation $\gs(n)$ at the  integral symbol is used for the ordered
string of variables  $\gs_1,\gs_2,\ldots \gs_n$.

Introducing additional integration parameters $\gs'{}^{ij}$ and
  new measure factors
\be
\label{muprime}
\mu'{}^i(\gt,\gs,\gs')=\mu^i(\gt,\gs)\,\prod_{j=0}^l d\gs'{}^{ij} \gd(\gs'{}^{ij}-a^{ij}(\gs)) \,  ,\ee
 one brings  $\Go^i_\ga$ to the form
\eq{Omega=}. Note that in \cite{Vasiliev:2023yzx}  it was proposed to consider polyhedra as integration domains, while
Eq.~\eq{muprime} provides some    variety embedded into a  polyhedron.
In this paper it is more convenient to  use
\eq{fetabeta} with $\Go$, $\bar \Go$ \eq{ Omegai} with some polyhedra as integration domains.

Another difference compared to the approach of \cite{Vasiliev:2023yzx} is that in this paper
we discard the weak terms, reconstructing the final results from the
compatibility conditions. Though we agree with the idea of \cite{Vasiliev:2023yzx} that
 it is useful to keep the weak terms inducing
  non-zero contribution at the further stages of the computations preserving the form
of the Ansatz we find it simpler to discard the weak terms in this paper
since our aim  is just to illustrate how moderately non-local vertex can be obtained in the mixed sector without going too much into
the computation details.

Due to the relation (\ref{xyy}) between space-time and spinor derivatives,
 to check whether $F^{i}$ \eq{fetabeta} with $G_l(g)$  \eq{G34}
is spin-local or not  one has  to analyse coefficients in front  of the terms bilinear in spinor derivatives
 $p_i{}_\ga p_j{}^\ga $ and $\bp_i{}_\pa \bp_j{}^\pa$  with respect to arguments
 of the zero-forms $C_i$ (\ie with with $i,j>0$) in the exponents of $\Ex  $ and $\bEx $ in \eq{EEE}.

\section{Moderate spin-non-locality}
 \subsection{Spin-locality and moderate spin-non-locality  }
\label{LOCALITY}
To check whether  $F^i  $ \eq{fetabeta}
    is spin-local or not
  we consider the coefficients  in front  of   $p_k{}_\ga p_j{}^\ga $ { and}
  $\bp_k{}_\pa \bp_j{}^\pa$   in the exponents of    $\EEE(\wmv^i |\bwmv^i \,)$, which yield, schematically,
\bee\label{regbexp}
\ls&& \exp  i \big( \gt z_\ga y^\ga +\ldots \!
\!+\! \half  P^{kj}  p_k{}^\ga p_j{}_\ga +
 {\bgt} \bz_\pa \by^\pa \!+\ldots
\!+\! \half  \bar{P}^{kj}  \bp_k{}^\pa \bp_j{}_\pa \big).\eee
 By the $Z$-dominance Lemma of \cite{Gelfond:2018vmi} (see also \cite{Didenko:2022eso}),
only  the coefficients   at $\gt=\bgt=0 $ matter.

$\bullet$ {\bf Spin-locality}\\
Spin-locality demands \cite{Gelfond:2018vmi} the condition

\be\label{Ploc}
P^{ij} \bar P^{ij}|_{\gt=\bgt=0} = 0
\qquad  \forall\, i,j>0\, . \ee

$\bullet$ {\bf Spin-non-locality}\\
Violation of this condition for at least one pair of $i,j>0$
implies spin-non-locality,
\be\label{Pnloc} \exists\, i\,,j >0 \qquad
P^{ij} \bar P^{ij}|_{\gt=\bgt=0} \ne 0
\,. \ee

$\bullet$ {\bf  Moderate spin-non-locality}\\
Here we  introduce the concept of {\it moderate spin-non-locality (MNL)} with the
coefficients $P^{ij}$ and $\bar P^{ij}$ obeying  the conditions
\be\label{nerav}
(|P^{ij}| + |\bar P^{ij}|)|_{\gt=\bgt=0} \le  1
\qquad  \forall i,j>0\, . \ee

Note that, generally,  spin-local vertices automatically obey the MNL conditions.
For instance, the  lower-order computations for  vertices bilinear  in $C$
in the (anti)holomorphic  sectors \cite{Vasiliev:2016xui,Gelfond:2017wrh,Didenko:2018fgx,Didenko:2019xzz}
imply that they satisfy both condition
\eq{Ploc} and \eq{nerav},
\be\label{ravnerav}
P^{12}\bar P^{12}|_{\gt=\bgt=0}=0\q (|P^{12}|+ |\bar P^{12}|)|_{\gt=\bgt=0} = 1 \,.\ee

It is not hard to find $ P^{ij}$   and $\bar P^{ij}$ \eq{regbexp} for the  Ansatz
 \eq{fetabeta}. For instance, for
   \be
 \wmv {}^\ga |_{\tau=0} = -\big(
a^0 p_0+a^1p_1   +\ldots + a^n    p_n{}
  \big) ^\ga \q
 \bwmv ^\pa|_{\bar\tau=0} = -(  \bar a^0\bp_0+ \bar a^1\bp_1 + \ldots +  \bar a^n \bp_n{}
 )^\pa\,
 \ee
one obtains
  \bee\label{PijAnzac}  P^{ij}|_{\gt=\bgt=0}=  a^i -a^j +1, \quad
  \bar P^{ij}|_{\gt=\bgt=0}=\bar  a^i - \bar a^j +1 \quad
\forall \le i<j\le n\,.\eee

 Note that the star product $C(y,\by)*C(y,\by)$ (\ref{star}) yields $|P^{12}|+|\bar P^{12}|=2$.

\subsection{Moderate non-locality compatible interpolating homotopy  }
\label{SimlpeHom}

Consider  equation  of the form
\be
\label{dab}
\dr A=F \,\q \dr F=0 .
\ee
Let $F$  be (i) of the form \eq{fetabeta} and (ii)  MNL.
To proceed we need a scheme allowing to solve (\ref{dab}) within the same class.
This is achieved by a MNL compatible interpolating homotopy (IH) introduced in this section.

Let  two expressions $F^a$ and $F^b$ be of the form \eq {fetabeta} and
\be
 \label{RaznE}
F^a+F^b=  \eta\bar\eta  \int\limits_{\gt \bgt \gs(n)  }
 \Big\{\mu^a( \gt,\bgt,\gs  )\EEE(\wmv^a |\bwmv^a\,)-\mu^b( \gt,\bgt,\gs  ) \EEE(\wmv^b\, |\bwmv^b)
   \Big\}G_l(g,K)   \,.  \ee
 Suppose that there exists   such a measure $\mu (\gn, \gt,\bgt,\gs  )$ depending on an
 additional parameter $\nu$, that
   $$\mu (\gn, \gt,\bgt,\gs  )|_{\gn=1}= \mu^b( \gt,\bgt,\gs  )\q
   \mu (\gn, \gt,\bgt,\gs  )|_{\gn=0}=\mu^a( \gt,\bgt,\gs  ).$$

 Since \be\dr[\theta(\gn )\theta(1-\gn )]
= d  \gn  ( \gd( \gn )-  \gd(1-\gn ))
 \,, \ee
  \bee
&&
 \label{RaznE=}
F^{a}+F^{b}=     \eta\bar\eta\int\limits_{   \gt \bgt \gs(n) \gn  } { } \,
 \mu'(\gn, \gt,\bgt,\gs  )
  \EEE( \wmv^\gn  |   \bwmv^\gn )
 G_l(g,K)\,,\eee
  where \be\label{muI}\mu'(\gn, \gt,\bgt,\gs  )=
 \dr[\theta(\gn )\theta(1-\gn )] \mu(\gn,  \gt,\bgt,\gs  )\,,\ee
 \be \label{OmegaIH}
 \wmv^\gn=  \gn  \wmv^b + (1-\gn) \wmv^a\,\q \bwmv^\gn=\gn  \bwmv^b + (1-\gn)\bwmv^a\,. \ee
 In these terms, the total differential $\dr$     \eq{fulldif}
acquires the form
\be\label{VasDifnu}
\dr = \theta^\ga  \f{\p}{\p z^\ga }+ \bar\theta^\pa  \f{\p}{\p \bz^\pa } +d\tau\f{\p}{\p \tau}+d\bgt \f{\p}{\p \bgt}+ d\gs_i \f{\p}{\p \gs_i}+
 d\nu \f{\p}{\p \nu }\,.
 \ee
Since the property \eq{dGOGO=0} is
still  true,
\be\label{dGOGO=0nI}
\dr\Big[ \EEE( \wmv^\gn  |   \bwmv^\gn )
\Big]=\dr\Big[(\dr\wmv^\gn)^2(\dr\bwmv^\gn)^2\mathcal{E}( \wmv^\gn)\bar {\mathcal{E}}(  \bwmv^\gn )
\Big]=0\,,
\ee
(\ref{muI}) allows us to represent $F^{a,b}$ \eq{RaznE=} in the form
 \bee
&&
 \label{RaznE==}
F^{a}+F^{b}=
 \dr G^{a,b} \,+ F^{a,b} \q\\
&& \label{dG12} G^{a,b}=   \eta\bar\eta  \int\limits_{  \gt \bgt \gs(n)  \gn } { } \,\mu'(\gn,  \gt,\bgt,\gs  )
  \EEE( \wmv^\gn  |   \bwmv^\gn )
 G_l(g,K)\,\q
 \\ \label{leftovF12}&&
F^{a,b}= -\eta\bar\eta\int\limits_{    \gt \bgt \gs(n) \gn   } { } \,
 \theta(\gn )\theta(1-\gn ) \dr\Big[ \mu( \gn, \gt,\bgt,\gs  )\Big]
 \EEE( \wmv^\gn  |   \bwmv^\gn )
 G_l(g,K)\,.\eee
 If  $F^a$ and $F^b$ \eq{RaznE}
 are MNL,
\ie
 $P^a{}^{ ij}$ and $\bar P^a{}^{ ij} $ of   $\EEE(\wmv^a\,|\bwmv^a\,)$
as well as  $P^b{}^{ij}$ and $\bar P^b{}^{ ij} $ of  $\EEE(\wmv^b\,|\bwmv^b\,)$   obey \eq{nerav},
 \be\label{ner12}
 (|P^a{}^{ij}|+|\bar P^a{}^{ij}|)|_{\gt=\bgt=0}\le1\q (|P^b{}^{ij}|+|\bar P^b{}^{ij}|)|_{\gt=\bgt=0}\le1\q i,j>0\,,\ee
 this is also true for $P^\gn{}^{ij}$ and $\bar P^\gn{}^{ij} $  of
$\EEE( \wmv^\gn|    \bwmv^\gn)$ with $ \wmv^\gn\,, \bwmv^\gn$\eq{OmegaIH}
   for any   $\gn\in[0,1] $.
Indeed, according to \eq{OmegaIH},  \eq{ExpO}, \eq{bExp} and  \eq{EEE}
  \bee \label{EEEI}
\EEE(\wmv^\gn,\bwmv^\gn)=\dr(\wmv^\gn)^2
\dr  (\bwmv^\gn)^2
\exp i \Big[\gn\Big\{ \Go^a_\gb(y^\gb +p^\gb_+ )  -\!\sum_{l\geq j>i\geq 1} p_{i\gb}
p_j^\gb -\!\sum_{l\geq j \geq 1} s_j p_{0\gb} p_j^\gb
\Big\}
\\ \nn +(1-\gn)\Big\{ \Go^b_\gb(y^\gb +p^\gb_+ )  -\!\sum_{l\geq j>i\geq 1} p_{i\gb}
p_j^\gb -\!\sum_{l\geq j \geq 1} s_j p_{0\gb} p_j^\gb
\Big\}\Big]\\ \nn
\times\exp i \Big[\gn\Big \{ \bar{\Go}^a_\pb(\by +\bp _+  )^\pb  -\!\sum_{l\geq j>i\geq 1} \bp_{i\pb}
\bp_j^\pb -\!\sum_{l\geq j \geq 1} \bar s_j \bp_{0\pb} \bp_j^\pb\Big\}
 \\ \nn
+(1-\gn)\Big\{\bar{\Go}^b_\pb(\by +\bp _+  )^\pb  -\!\sum_{l\geq j>i\geq 1} \bp_{i\pb}
\bp_j^\pb -\!\sum_{l\geq j \geq 1} \bar s_j \bp_{0\pb} \bp_j^\pb\Big \}
 \Big]
 \,.
\eee
Rewriting exponents in the form   \eq{regbexp},
one obtains  \be \label{Pijnu} P^\gn{}^{ij}= \gn P^a{}^{ij}+(1-\gn) P^b{}^{ij}\q
 \bar  P^\gn{}^{ij}= \gn \bar P^a{}^{ij}+(1-\gn) \bar P^b{}^{ij}\,.\ee
Since $ \gn\in [0,1]$, \eq{ner12} and \eq{Pijnu} imply
$(|P^\gn{}^{ij}|+|\bar  P^\gn{}^{ij}|)|_{\gt=\bgt=0}\le 1.$ The essence of the idea is
that if the  coefficients  $a^i{}^j(\gs)$  for any $i,j$ on the
\rhs  of \eq{ Omegai}
satisfy
\be\label{ineqcoef} |a^i{}^j(\gs)|\le1
  \ee
then\be\label{ineqcoefnu}
   |\gn a^a{}^j(\gs)+(1-\gn) a^b{}^j(\gs)|\le1
 \ee
 as well.
In the sequel it will be shown, that inequality \eq{ineqcoef}  holds true
for a     set of functions $\wmv$, $\bwmv$
under consideration, thus   forming a   convex set.

    Picking up an appropriate  pair   $F^{a}$ and $F^{b}$   on the \rhs of \eq{dab}
 we apply IH to single out the corresponding
$\dr$-exact part  setting $A=G^{a,b}+ A'$ we are left with the equation
\be
\label{dabab}
\dr A'=\sum_{i\ne a,b} F^i + F^{a,b}\,,
\ee
with  $F^{a,b}$ \eq{leftovF12}. The \rhs of \eq{dabab} is evidently (i) $\dr$-closed,  (ii)
 of the form \eq{fetabeta} and (iii)  MNL as the \rhs of \eq{dab}.

 To arrive at the final result we repeat this procedure as many times as needed for the
leftover MNL terms  until   all
of them cancel.
 Note that, at every step,  the choice of a proper pair is to large extent   ambiguous
and it is not {\it a priory}  guaranteed that the process  ends at some stage. For instance, the choice
$\mu^b=0$ can unlikely yield  a reasonable result.

 Nevertheless, for some reason to be  better understood, it works. Let us stress that in this paper we manage to choose all appropriate pairs of  the \rhss under consideration  with the
same measure factors $\mu^a=\mu^b$, that simplifies the calculations making $\mu (\gn, \gt,\bgt,\gs  )$ $\nu$-independent.

 This   interpolating homotopy approach      underlies the construction of MNL solutions.
Specifically, it  is used below to solve for $S_2$ the following consequences of
\eq{HS2}
\be\label{dS2mixeq}
2i\dr  S_2^{\eta\bar\eta}  =-\left\{i\bar\eta B_2^{\eta}*
\bar\gamma+i\eta  B_2^{\bar\eta}* \gamma -\{ S_1^{\bar \eta}\,
S_1^{\eta}\}_* \right\}
\ee    in such a way that the
\rhss of the following  consequences of
\eq{HS1},  \eq{HS3}
\bee\label{dW2=}
2i\dr   {W}^{\eta \bar\eta}_2 &=& \left\{\dr_x S_1^\eta+\dr_x S_1^{\bar\eta}
+\{W_1^\eta\,, S_1^{\bar\eta}\}_\ast+\{W_1^{\bar\eta}\,,S_1^{ \eta} \}_\ast
+\dr_x S_2^{\eta\eta}+\{\omega\,, S_2^{\eta\bar\eta}\}_*
\right\}\,,
\\ \label{DB3}2i \Dt   B_3^{\eta\bar \eta}&=& \left\{[S_1^{\eta } ,B_2^{ \bar \eta} ]_*+
  [S_1^{ \bar \eta} ,B_2^{\eta } ]_*
+ [S^{\eta\bar \eta}_2,C]_*\right\}  \eee
as well as $[\Dt W^{\eta \bar\eta}_2,C]_*$ be MNL.

This  allows us to find by IH such $B_3^{\eta\bar \eta}$
that the \rhs of
\bee \label{UpsfullwCCC}
\Upsilon^{\eta\bar\eta}(\go,\!C,\!C,\!C)= &&
\ls-[W^{\eta\bar \eta}_2 \,,C]_*
-[ {W}^{\bar \eta}_1\,, B^{  \eta}_2]_*-[{W}^{  \eta}_1\,,   B^{\bar \eta}_2]_*
- \dr_x B^{\eta \bar \eta}_3
\\ \nn &&\ls-[\go\,,B^{\eta \bar \eta}_3]_*
-\dr_x B^{  \eta}_2 (\Upsilon^{\bar \eta}(\go,C,C ))-\dr_x B^{\bar \eta}_2 (\Upsilon^{ \eta}(\go,C,C ))\,
\eee
  in its turn becomes MNL, allowing  to eliminate step by step manifest  $Z$-dependence
  using IH. Namely, choosing an appropriate  pair  from the \rhs of \eq{UpsfullwCCC}
 we apply IH to
 drop the $\dr$-exact part since it does not contribute in the $Z,\,dZ$-independent sector.
 Then  this procedure is repeated as many times as needed
  until   all leftover MNL terms cancel  except for the
 cohomological terms producing MNL physical vertices
(see Section \ref{MNLvert}).

 Note that the interpolating homotopy can be treated as certain generalization of the general homotopy of \cite{Vasiliev:2023yzx}.

\section{ Moderately non-local  $ B_3^{\eta\bar \eta }$}
\label{MNLB3}

To compute the  MNL form of  $\Upsilon^{\eta\bar\eta}({\go, C,C,C}) $ vertex
we have to find a MNL  $B_3$.  This is the aim of this section.
In the sequel we use notations of \cite{Vasiliev:2023yzx}
\bee&&\label{nabla}
 \Box(\gt,\bgt)=l(\gt)l(\bgt)\q l(\gn)=\theta(\nu)\theta(1-\nu)\q
  \tr(\nu)=d \gn \gd(\gn )\,,
\\
&& \nn
  \nabla(\ga(n)):=\prod^n_{i=1} \theta(\ga_i)\tr\left(1-\sum^n_{i=1} \ga_i\right)\,.
 \eee

Equation for $B^{\eta\bar\eta}_3$  in the mixed sector resulting from \eqref{HS5}
has the form \eq{DB3}. To obtain MNL $B_3$ we need  the \rhs of \eq{DB3} to
be of that class.
 Straightforwardly, using   $S_{1,2}$ and $B_2$ of \cite{Didenko:2019xzz},
 one can  make sure that this is true for $[S_1^{\eta } ,B_2^{ \bar \eta} ]_*+
  [S_1^{ \bar \eta} ,B_2^{\eta } ]_*$, while  $[ S^{\eta\bar\eta}_2 ,C]_*$ is not MNL.
The key observation of this paper is that, as we show now, there exists an alternative $S^{\eta\bar\eta}_2$ such that $[ S^{\eta\bar\eta}_2 ,C]_*$ is  MNL.

 \subsection{$ S_2 ^{\eta\bar\eta} $ }
\label{S2etabareta}

  $\dr S_2^{\eta\bar \eta}$ is determined by \eq{dS2mixeq}.
 One can  make sure straightforwardly that
 $[\dr S^{\eta\bar \eta}_2,C]_*$ is both spin-local and MNL.
The problem is that all spin-local terms of $[\dr S^{\eta\bar \eta}_2,C]_*$ have different structure and it is not clear how
to find such a solution for $S^{\eta\bar \eta}_2$  that $[S_2^{\eta\bar \eta}\,,C]_*$ be spin-local.
However, since $[\dr S^{\eta\bar \eta}_2,C]_*$ is MNL,  the interpolating homotopy of Section \ref{SimlpeHom} allows us to
find  such $ S^{\eta\bar \eta}_2 $  that $[S_2^{\eta\bar \eta}\,,C]_*$ is
 MNL as well.

 Indeed, one can see that
\bee\label{S1S1}
 -\{ S_1^{\bar \eta}\,,
S_1^{\eta}\}_* &=& \!\!\!  \ff{  \eta\bar{\eta}}{4}
 \!\!\int\limits_{\gt \bgt  }
 \Box(\gt,\bgt)
    [\EEE(\wmv^1 |\bwmv^1)-\EEE(\wmv^2 |\bwmv^2)]
 CC  k\bar{k}  ,\qquad
\\\label{10wS1S11}
 \wmv^1_{ \,\ga}&=& \gt   z_ \ga- (1-\gt  )[  -  p _1{}     ]_\ga   \q
  \bwmv^1_{ \,\pa}= \bgt   \bz_\pa -(1-\bgt  )[  \bp_2{}   ]_\pa\,,
\\\label{10wS1S12}
 \wmv^2_{ \,\ga}&=& \gt   z_ \ga- (1-\gt  )[       p_2{} ]_\ga   \q
  \bwmv^2_{ \,\pa}= \bgt   \bz_\pa -(1-\bgt  )[-   \bp_1{}  ]_\pa\,.\eee
 Applying IH to the \rhs of \eq{S1S1}
one finds   $S_2$  in the  form \eq{fetabeta}.
Namely, one can see that
\bee\label{S1S1=}
-\{ S_1^{\bar \eta}\,,
S_1^{\eta}\}_* &=& \!\!\! \dr\Big[ \ff{ \eta\bar{\eta}}{4}
 \!\!\int\limits_{\gt \bgt \gs( 2)} \nabla(\gs(2))\Box(\gt,\bgt)
    \EEE(\wmv |\bwmv)\Big]
 CC  k\bar{k}\\ \label{bpS1S1}
&&- \ff{ \eta\bar{\eta}}{4}
 \!\!\int\limits_{\gt \bgt \gs( 2)}  \dr[\Box(\gt,\bgt) ]  \nabla(\gs(2))
    \EEE(\wmv |\bwmv)
 CC  k\bar{k}   ,\qquad
\eee
\bee\label{10wS2}
 \wmv_{ \,\ga}= \gt   z_ \ga- (1-\gt  )[  -\gs_1  p _1{}+    \gs_2    p_2{} ]_\ga   \q
  \bwmv_{ \,\pa}&=& \bgt   \bz_\pa -(1-\bgt  )[- \gs_2 \bp_1{} + \gs_1 \bp_2{}   ]_\pa\,.
\eee
Differentiation of $\Box(\gt,\bgt)$   in  \eq{bpS1S1} yields
\bee \label{B2gga}&&  \ls \ff{\eta\bar{\eta}}{4}
 \!\!\!\!\int\limits_{\gt \bgt \gs( 2)} \!\!\! \!\nabla(\gs(2))
   \Big[ \tr(1-\gt)   l(\bgt) \EEE(\wmv'  |\bwmv)+ \tr(1-\bgt)   l(\gt)
   \EEE(\wmv  |\bwmv')\Big]
 CC  k\bar{k}\,,  \eee
where
$ \wmv'_{ \,\ga}=     z_ \ga    \,,
  \bwmv'_{ \,\pa}=    \bz_\pa  \,,$
while  the (weak) terms with $\tr(\gt)$ or  $\tr(\bgt)$ vanish  because, e.g.,
 $$
 \tr(\gt) \tr(1-\gs_1-\gs_2)\dr \wmv_{ \,\ga}\dr \wmv^{ \,\ga}\sim
d\gt  d (\gs_1+\gs_2) d  \gs_2 d \gs_1   \gd(\gt)\gd(1-\gs_1-\gs_2)p_1{}_\ga p_2{}^\ga =0\,.
 $$
As a result, \eq{B2gga} just equals to $-i\bar\eta B_2^{\eta}*
\bar\gamma-i\eta  B_2^{\bar\eta}* \gamma $
while Eq.~\eq{dS2mixeq} acquires the form
\bee\label{dS2mixeq=}
2i\dr  S_2^{\eta\bar\eta}   &=& \!\!\!- \dr\Big[ \ff{ \eta\bar{\eta}}{4}
 \!\!\int\limits_{\gt \bgt \gs( 2)}   \Box(\gt,\bgt)  \nabla(\gs(2))
    \EEE(\wmv |\bwmv)\Big] CC  k\bar{k}
\eee
allowing to set
    \bee\label{10intsS2}
 S_2^{\eta\bar \eta}\, &=& \!\!\!   \ff{i \eta\bar{\eta}}{8}
 \!\!\int\limits_{\gt \bgt \gs( 2)}  \Box(\gt,\bgt) \nabla(\gs(2))
    \EEE(\wmv |\bwmv)
 CC  k\bar{k}  \,.
\eee

By construction, $S_2^{\eta\bar \eta}$ \eq{10intsS2} is spin-local, while  $[S_2^{\eta\bar \eta}\,,C]_*$ is MNL. Indeed, consider
 for instance  the exponent of    $S^{ \eta\bar\eta}_2*C$ in the form
  \eq{regbexp}, \ie
\be\label{regexpS2}
  \exp  i \big( ...\!+\! \half  P^{ij}  p_i{}^\ga p_j{}_\ga
  \!+\! \half  \bar{P}^{ij}  \bp_i{}^\pa \bp_j{}_\pa \big).\ee
Eq.~\eq{10intsS2} straightforwardly    yields
by virtue of  Eq.~\eq{10wS2}
\bee\label{wS2CP}
 P^{12}|_{\gt=\bgt=0}= 0\q   P^{13}|_{\gt=\bgt=0}= \gs_1 \q   P^{23}|_{\gt=\bgt=0}=  -\gs_2\,,
\\ \nn
 \bar P^{12}|_{\gt=\bgt=0}= 0\q \bar P^{13}|_{\gt=\bgt=0}= \gs_2 \q \bar P^{23}|_{\gt=\bgt=0}=  -\gs_1\,.
\eee
Thanks to $\Delta(1-\gs_1-\gs_2)$ on the
\rhs of Eq.~\eq{10intsS2}  inequalities \eq{nerav} hold true.

   \subsection{$\dr B^{\eta\bar\eta}_3$ }

Substituting $S_1$ ,$W_1$, $B_2$    \eq{S1withw}-\eq{bB2f}, $S_2$ \eq{10intsS2}
we obtain using \eq{nabla}
  \be
 \label{10S2C}
 \ff{1}{2i} S_2{}^{\eta\bar\eta}*C{}  = \!-\!
\ff{ \eta\bar{\eta}}{16}
\!\! \int\limits_{\gt \bgt \gs( 2)}\!\! \Box(\gt,\bgt)\nabla(\gs(2))
      \EEE(\wmv\, |\bwmv)
 CCC    k\bar{k}  \,,
 \ee
 \be
\wmv^\ga= \gt   z{}^\ga\!-\! (1\!-\!\gt  )[  \!-\!\gs_1  (p _1{}+ p_2) +  p_3{}+ p_2  ]^\ga  \q
\bwmv^\pa= \bgt   \bz{}^\pa \!-\!(1\!-\!\bgt  )[\!-\! \gs_2 (\bp_1{}+\bp_2)    + \bp_3{}+\bp_2 ]^\pa\,,
\ee
    \be
 \label{10CS2C}
    \!-\!\ff{1}{2i}C*S_2{}^{\eta\bar\eta}  =
\ff{ \eta\bar{\eta}}{16}\!\! \int\limits_{\gt \bgt \gs( 2)}\!\! \Box(\gt,\bgt)\nabla(\gs(2))
     \EEE(\wmv\, |\bwmv) CCC k\bar{k} \,,
     \ee
     \be
\label{wCS2}
\wmv^\ga= \gt   z{}^\ga\!-\! (1\!-\!\gt  )[ \!-\!p_1  \!-\! p _2{} +    \gs_2   ( p_3+p_2){}  ]^\ga  \q
\bwmv^\pa= \bgt   \bz{}^\pa \!-\!(1\!-\!\bgt  )[\!-\!\bp_1{}\!-\!   \bp_2{} + \gs_1 (\bp_3+\bp_2){}  ]^\pa\,,
\ee
 \be
 \label{10barb2S1}
   \!-\!\ff{1}{2i}B^{ \eta}_2*\bar S^{\bar\eta}_1  =
\ff{ \eta\bar{\eta}}{16}\!\! \int\limits_{\gt \bgt \gs( 2)}\!\! \Box(\gt,\bgt)\nabla(\gs(2))
 \EEE(\wmv\, |\bwmv) CCC k\bar{k}\,,
 \ee
 \be
\label{wbSB2}
\wmv{}^\ga= \gt   z^\ga\!-\!   (1\!-\!\gt  )[
 \!-\! \gs_1 (  p_1+  p_2)+ p_2+   p_3 ]^\ga  \q
\bwmv^\pa = \bgt   \bz^\pa \!-\!(1\!-\!\bgt  )[ \!-\!\bp_1\!-\!   \bp_2 ]^\pa   \,,
\ee
   \be
 \label{barS1B2} \ff{1}{2i} \bar S^{\bar\eta}_1 * B^{ \eta}_2\,          = \!-\!
\ff{ \eta\bar{\eta}}{16}\!\! \int\limits_{\gt \bgt \gs( 2)}\!\! \Box(\gt,\bgt)\nabla(\gs(2))
 \EEE(\wmv\, |\bwmv) CCC k\bar{k} \,,
 \ee
 \be
\label{wB2bS}\wmv^\ga= \gt   z^\ga\!-\!   (1\!-\!\gt  )[
 \!-\!  p_1 \!-\!  p_2+   \gs_2 (p_3+ p_2) ]^\ga  \q
  \bwmv{}^\pa = \bgt   \bz^\pa \!-\!(1\!-\!\bgt  )[ \bp_2+ \bp_3]^\pa   \,,
\ee
 \be\label{barB2S1}
 \!-\!\ff{1}{2i}\bar B^{\bar\eta}_2*S^{ \eta}_1
  =  \!-\!
\ff{ \eta\bar{\eta}}{16}\!\! \int\limits_{\gt \bgt \gs( 2)}\!\! \Box(\gt,\bgt)\nabla(\gs(2))
   \EEE(\wmv\, |\bwmv) CCC k\bar{k}\,,
   \ee
   \be
\label{SbB}\wmv^\ga= \gt   z^\ga\!-\!   (1\!-\!\gt  )[\!-\!  p_1  \!-\!   p_2   ]^\ga  \q
\bwmv{}^\pa = \bgt   \bz^\pa \!-\!(1\!-\!\bgt  )[
  \!-\! \gs_1( \bp_1+ \bp_2)+    \bp_2+  \bp_3 ]^\pa   \,,
\ee
\be
 \label{10S1barB2}
 \ff{1}{2i}S^{ \eta}_1  *\bar B^{\bar\eta}_2   =
\ff{ \eta\bar{\eta}}{16}\!\! \int\limits_{\gt \bgt \gs( 2)}\!\! \Box(\gt,\bgt)\nabla(\gs(2))   \EEE(\wmv\, |\bwmv) CCC k\bar{k}\,,
\ee
\be
\wmv^\ga= \gt   z^\ga\!-\!   (1\!-\!\gt  )[ p_2+  p_3  ]^\ga  \q
\bwmv{}^\pa = \bgt   \bz^\pa \!-\!(1\!-\!\bgt  )[
  \!-\! \bp_1\!-\!    \bp_2+ \gs_2 ( \bp_3+\bp_2) ]^\pa  \,.
\ee

As mentioned in Section \ref{S2etabareta},
 the \rhss of Eqs.~\eq{10S2C} and \eq{10CS2C} are MNL.
Straightforwardly one can check that   the \rhss of Eqs.~\eq {10barb2S1}, \eq{barS1B2},
  \eq{barB2S1} and \eq{10S1barB2}   are
 also  MNL.
Indeed,
 consider for instance the \rhs of \eq{10S1barB2}. According to
  Eqs.~\eq{EEE}-\eq{ Omegai}  the exponent is
\bee \label{provdB3}
 \exp i \big( (\gt   z -   (1-\gt  )[ p_2+  p_3  ]  )_\gb(y +p_+ )^\gb  -\!\sum_{3\geq j>i\geq 1} p_{i\gb}
p_j^\gb
\big)\times
\\ \nn
\exp i \big ((\bgt   \bz  -(1-\bgt  )[
  - \bp_1-    \bp_2+ \gs_2 ( \bp_3+\bp_2) ])_\pb(\by +\bp _+  )^\pb  -\!\sum_{3\geq j>i\geq 1}
 \bp_{i\pb}
\bp_j^\pb  \big).
\eee
Discarding the $\gt$, $ \bgt$, $y$ and $\bar y$-dependent terms  one is left with
 \bee \nn i \big(...
-  p_2{}_\gb p_3{}^\gb
 -  \gs_2   \bp_3{}_\pb \bp _1  {}^\pb +  (1-\gs_2)\bp_2{}_\pb  \bp_1  ^\pb
   \big).
\eee
Since the coefficients in front of $ p_i{}_\gb   p_j  ^\gb$ and $\bp_i{}_\pb  \bp_j  ^\pb$ satisfy  inequalities \eq{nerav} $S^{ \eta}_1  *\bar B^{\bar\eta}_2$ is MNL. Note that  it is also  spin-local.

That the  \rhss of  Eqs.~\eq {10barb2S1}, \eq{barS1B2} and
  \eq{barB2S1}  are MNL  can be checked analogously.
 Once $\dr B^{ \eta\bar\eta}_3$ is shown to be MNL one can look for
 MNL $B_3$ applying IH.

\subsection{Solving for  moderately non-local $B^{\eta\bar\eta}_3$}
\label{solvB3}

Applying IH
to the sum of \eq{10barb2S1} and \eq{barB2S1} and then  of
\eq{barS1B2} and \eq{10S1barB2}, using \eq{RaznE==}
one can see that the
terms \eq{10S2C} and \eq{10CS2C} cancel out and  \eq{DB3} yields
 using notation \eq{nabla}
  \bee\ls\ls  &&\Dt   B_3^{\eta\bar \eta}=
 \label{dB3bulk}\ff{\eta\bar{\eta}}{16}
\ls    \int\limits_{\gn(2)\gt \bgt \gs( 2)}
\ls\,\Big\{
   \Dt \big[  \Box(\gt,\bgt)  \nabla(\gn(2)) \nabla(\gs(2))
   \big]
\\  &&
  \label{B3dBox}
  - \Dt[\Box(\gt,\bgt)] \nabla(\gn(2)) \nabla(\gs(2))\Big\}
 \Big[ \EEE(\wmv_1\, |\bwmv_1)-\EEE(\wmv_2\, |\bwmv_2)\Big] CCC k\bar{k}\,,
 \eee
where
  \bee\label{21w1B3}
 &&\wmv_1^{ \ga}:= \gt   z{}^\ga- (1-\gt  )[  - \gn_2  (p _1{}+ p_2)^\ga+ (1-\gn_2\gs_1)(p_3{}+ p_2)^\ga]  \q
 \\ \nn
&&\bwmv_1^{ \pa}:= \bgt   \bz{}^\pa -(1-\bgt  )[-  \gn_1 (\bp_1{}+\bp_2)^\pa
+(1-\gn_1\gs_1)(\bp_3{}+\bp_2)^\pa]\,,
  \\\label{21w2B3}
 &&\wmv_2^{ \ga}:= \gt   z{}^\ga- (1-\gt  )[  -( 1-\gn_2\gs_2) (p _1{}+ p_2)^\ga+  \gn_2 (p_3{}+ p_2)^\ga]  \q
 \\ \nn
&&\bwmv_2^{ \pa}:= \bgt   \bz{}^\pa -(1-\bgt  )[ -(1-\gn_1\gs_2) (\bp_1{}+\bp_2)^\pa
  + \gn_1(\bp_3{}+\bp_2)^\pa]\,. \eee
Since it is shown that Eqs.~\eq {10barb2S1}, \eq{barS1B2},
  \eq{barB2S1} and \eq{10S1barB2}
  are MNL,  by the reasoning of Section \ref{LOCALITY} all   terms on the \rhss of
  \eq{dB3bulk} and \eq{B3dBox}  are MNL as well.

 Eq.~\eq{dB3bulk} determines a part of $B_3^{\eta\bar\eta}$ with the integrand containing
  $  \Box(\gt,\bgt)$ without derivatives. Following \cite{Vasiliev:2023yzx}, such terms
will be referred to as 'bulk' in contrast to thouse with $  \dr \Box(\gt,\bgt)$  referred to as 'boundary',
 \be   \Dt  \Box(\gt,\bgt)=[ \tr(\gt)+\tr (1-\gt)]l(\bgt)+c.c.
 \ee

The  terms  proportional to $\tr(1-\gt)$
or  $\tr(1-\bgt)$  do not contribute  to \eq{B3dBox} (are weakly zero
in terminology of \cite{Vasiliev:2023yzx})  because of the
  lack of differentials.
Indeed, consider  for instance the $\Omega_1$-dependent term with
$\sim \tr(1-\bgt)$.
Due to \eq{EEE}
along with \eq{21w1B3}, \eq{nabla}
 it yields
  \bee\label{gt1=1}
&& \ldots d \bgt\gd(1-\bgt) l(\gt)  \nabla(\gn(2)) \nabla(\gs(2))
  (\dr\wmv^1)^2
( \bar  \theta_\pa \bar\theta^\pa)^2\E(\Go^1)  \bar\E(\bar\Go^1)\ldots
  \ldots\eee
  Since non-weak terms of $(\dr\wmv^1)^2$ must contain $\dr\gt$, modulo weak terms it equals to  \be\label{weak1}
    2d \gt
\Big\{   z{} +[  -( 1-\gn_2\gs_2) (p _1{}+ p_2) +  \gn_2 (p_3{}+ p_2)]\Big\}_\ga
\Big\{\gt   d z{} - (1-\gt  )[   d( \gn_2\gs_2) (p _1{}+ p_2) +  d\gn_2 (p_3{}+ p_2)]\Big\}^\ga.
\ee
To be non-weak it must contain a factor of $d \gs_2 d \gn_2$
 which is absent in \eq{weak1}.

 Hence
   non-zero 'boundary' terms  are those proportional to  either     $ \tr(\gt) $ or $ \tr(\bgt)$.
Firstly, consider the terms with $\tr(\bgt)$.
 To see, that the sum of such  terms   is $\Dt-$closed,
 it is useful to make the following change of variables:
\bee \label{ratioksibgt}\gn_1\gs_1:=\gx_1\q\gn_1\gs_2:=\gx_2\q\gn_2=\gx_3\q    \sum\gx_i=1
   \qquad\eee
 with $\wmv_1,\bwmv_1$ \eq{21w1B3}.
 To change   variables in the $\wmv_2,\bwmv_2$  part  \eq{21w2B3}
we use the following cyclic permutation of \eq{ratioksibgt}
\bee \label{ratioksibgt2}\gn_1\gs_1:=\gx_2
\q\gn_1\gs_2:=\gx_3
\q\gn_2=\gx_1\q    \sum\gx_i=1
  \,.\eee
As a result, using notations \eq{nabla},
the   $\tr(\bar\gt)-$proportional part of  \eq{B3dBox}
 acquires the form
\bee  &&
 \label{B3dBoxbgt=0}
  \ls \ff{ \eta\bar{\eta}}{16} \!\!  \int\limits_{  \gt \bgt \gx(3)}
\ls
   \nabla(\gx(3))    \tr(\bgt)l(\gt)  \Big[ \EEE(\wmv_1\, |\bwmv_1)-\EEE(\wmv_2\, |\bwmv_2)\Big] CCC k\bar{k}\,
    , \eee
where \bee\label{21w1B3=}
 &&\wmv_1^{ \ga}:= \gt   z{}^\ga- (1-\gt  )[  - \gx_3  (p _1{}+ p_2)^\ga
 + \left(1-  {\gx_1\gx_3}({1-\gx_3})^{-1}\right)(p_3{}+ p_2)^\ga]  \q
 \\ \nn
&&\bwmv_1^{ \pa}:= - [ -(1-\gx_3) (\bp_1{}+\bp_2)
+ (1-\gx_1)(\bp_3{}+\bp_2)]^\pa\,,
\\\label{21w2B3=}
 &&\wmv_2^{ \ga}:= \gt   z{}^\ga- (1-\gt  )[    -\left(1-  {\gx_1\gx_3}({1-\gx_1})^{-1}\right) (p _1{}+ p_2)^\ga
 +  \gx_1 (p_3{}+ p_2)^\ga]  \q
 \\ \nn
&&\bwmv_2^{ \pa}:= -[ -(1-\gx_3) (\bp_1{}+\bp_2)
+ (1-\gx_1)(\bp_3{}+\bp_2) ]^\pa\,.
 \eee

Analogously, changing the variables in
 the   $\tr(\gt)$ part of  \eq{B3dBox}
 with   $\wmv_1\,,\bwmv_1$ \eq{21w1B3}
   \be \label{ratioksigt} \gn_2\gs_1:=\gx_1\q\gn_2\gs_2:=\gx_2\q\gn_1=\gx_3\q    \sum\gx_i=1
    \ee
and the  cyclically transformed change of variables $\gx_3\to\gx_1\to\gx_2\to\gx_3 $ in   $\wmv_2$ and $\bwmv_2$ \eq{21w2B3},
we obtain
 \bee  &&
 \label{B3dBoxgt=0}
  \ls \ff{ \eta\bar{\eta}}{16} \!\!  \int\limits_{  \gt \bgt \gx(3)}
\ls     \nabla(\gx(3))   \tr(\gt)l(\bgt)  \Big[ \EEE(\wmv_1\, |\bwmv_1)-\EEE(\wmv_2\, |\bwmv_2)\Big] CCC k\bar{k}
    ,\qquad  \eee
where
   \bee\label{31w1B3}
 &&\wmv_1^{ \ga}:= \gt   z{}^\ga- (1-\gt  )[  - ( 1-\gx_3)  (p _1{}+ p_2)^\ga+ (1-\gx_1)(p_3{}+ p_2)^\ga]  \q
 \\ \nn
&&\bwmv_1^{ \pa}:= \bgt   \bz{}^\pa -(1-\bgt  )[-  \gx_3 (\bp_1{}+\bp_2)^\pa
+(1-\gx_3\gx_1(1-\gx_3)^{-1})(\bp_3{}+\bp_2)^\pa]\,,
\\\label{31w2B3}
 &&\wmv_2^{ \ga}:= \gt   z{}^\ga- (1-\gt  )[  -( 1-\gx_3) (p _1{}+ p_2)^\ga+  ( 1-\gx_1) (p_3{}+ p_2)^\ga]  \q
 \\ \nn
&&\bwmv_2^{ \pa}:= \bgt   \bz{}^\pa -(1-\bgt  )[ -(1-\gx_1\gx_3(1-\gx_1)^{-1}) (\bp_1{}+\bp_2)^\pa
  + \gx_1(\bp_3{}+\bp_2)^\pa]\,
\,. \eee
   One can easily make sure that the  expressions \eq{B3dBoxbgt=0} and \eq{B3dBoxgt=0} are $\Dt-$closed.
   For instance, applying $\dr$ to \eq{B3dBoxgt=0}  one can see that the
   only potentially non-zero term is that
   with $ \tr(1-\bgt)$.
  However,  Eqs.~\eq{31w1B3}, \eq{31w2B3} yield
    $ \Big[ \EEE(\wmv_1\, |\bwmv_1)-\EEE(\wmv_2\, |\bwmv_2)\Big]\Big |_{\bgt=1}=0$.
    The case of \eq{B3dBoxbgt=0}  is analogous.

Application of  IH of Section \ref{SimlpeHom}
 to the MNL pairs  of \eq{B3dBoxbgt=0}    and   \eq{B3dBoxgt=0} brings
  the 'boundary' part of Eq.~\eq{B3bound} to the  form
\bee  &&
 \label{DtB3dBox=}
  \ff{ \eta\bar{\eta}}{16}  \Dt\Big\{  \int\limits_{\ga(2) \gt \bgt \gx(3)}
  \!\! \nabla(\ga(2))   \nabla(\gx(3))
 \Big[-\tr(\bgt) l(\gt)  \EEE(\wmv_3\, |\bwmv_3)
 +\tr(\gt) l(\bgt)  \EEE(\wmv_4\, |\bwmv_4)\Big] CCC k\bar{k}
\Big\}\, ,\qquad\rule{20pt}{0pt}  \eee where
  \bee\label{31w3B3===}
 &&\wmv_3^{ \ga}:= \gt   z{}^\ga- (1-\gt  )
 \left [    -\left \{ \ga_1\gx_3+\ga_2  (1-\gx_1\gx_3(1-\gx_1)^{-1})
  \right\} (p _1{}+ p_2)^\ga\right. \\ \nn&&
 \left.
 + \left\{\ga_1(1-\gx_3\gx_1(1-\gx_3)^{-1})+\ga_2\gx_1\right\}(p_3{}+ p_2)^\ga\right]  \q
 \\ \nn
&&\bwmv_3^{ \pa}:= -[ -(1-\gx_3) (\bp_1{}+\bp_2) + (1-\gx_1)(\bp_3{}+\bp_2) ]^\pa\,,
 \\ \label{31w4B3==}
 &&\wmv_4^{ \ga}:=-[  - ( 1-\gx_3)  (p _1{}+ p_2)^\ga+ (1-\gx_1)(p_3{}+ p_2)^\ga]  \q
 \\ \nn
&&\bwmv_4^{ \pa}:= \bgt   \bz{}^\pa -(1-\bgt  )[-
\Big\{ \ga_1\gx_3+\ga_2  (1-\gx_1\gx_3(1-\gx_1)^{-1})\Big\} (\bp_1{}+\bp_2)^\pa
 \\ &&\nn +\left\{\ga_1(1-\gx_3\gx_1(1-\gx_3)^{-1})+\ga_2\gx_1\right\}(\bp_3{}+\bp_2)^\pa]\,.
 \eee
  Eqs.~\eq{dB3bulk} and \eq{DtB3dBox=}
 yield the following final result for MNL $B_3$:
\bee &&  \label{B3bulk+bnd} \nn B_3^{\eta\bar \eta}=B^{\eta\bar \eta}_3|_{blk}
+B_3^{\eta\bar \eta}|_{bnd}\q
\\ &&
  \label{B3bulk}    B^{\eta\bar \eta}_3|_{blk}  =
   \ff{\eta\bar{\eta}}{16}\ls \int\limits_{\gn(2)\gt \bgt \gs( 2)}
  \!\! \Box(\gt,\bgt) \nabla(\gn(2))   \nabla(\gs(2))
  \Big[ \EEE(\wmv_1\, |\bwmv_1)
\!-\!\EEE(\wmv_2\, |\bwmv_2)\Big] CCC k\bar{k}   \rule{0pt}{20pt}
   \q
\\  \label{B3bound}
\ls  &&B_3^{\eta\bar \eta}|_{bnd}=\ff{ \eta\bar{\eta}}{16}\!\ls \int\limits_{\ga(2) \gt \bgt \gx(3)}\!\!
 \ls\nabla(\ga(2))   \nabla(\gx(3))%\\ \nn  &&\times
 \Big[ \tr(\bgt) l(\gt)  \EEE(\wmv_3\, |\bwmv_3)
\!-\!\tr(\gt)l(\bgt)  \EEE(\wmv_4\, |\bwmv_4)\Big] CCC k\bar{k}\rule{30pt}{0pt}
  \eee
with $\wmv_1, \bwmv_1$ \eq{21w1B3}, $\wmv_2, \bwmv_2$ \eq{21w2B3},
   $\wmv_3, \bwmv_3$ \eq{31w3B3===} and $\wmv_4, \bwmv_4$ \eq{31w4B3==}.
$B_3^{\eta\bar\eta}|_{blk}$  \eq{B3bulk} and $B_3^{\eta\bar\eta}|_{bnd}$ \eq{B3bound}  are
  MNL by construction.
 This  allows us to construct the MNL
vertex $\Upsilon^{\eta\bar\eta}({\go, C,C,C}) $.

   \section{Moderately non-local  vertex $\Upsilon^{\eta\bar\eta}({\go, C,C,C}) $  }
\label{MNLvert}

According to Eq.\eq{G34}
the   vertex $\Upsilon^{\eta\bar\eta}({\go, C,C,C}) $
   in the zero-form sector   can
 be represented in the form
\be \label{VertOrder}
\Upsilon^{\eta\bar\eta}(\go,C,C,C)=\Upsilon^{\eta\bar\eta}_{\go CCC}+\Upsilon^{\eta\bar\eta}_{C\go CC}
+\Upsilon^{\eta\bar\eta}_{CC\go C}+\Upsilon^{\eta\bar\eta}_{CCC\go }\,
\ee
with the subscripts   referring to the orderings of the product
factors.

 As a consequence of consistency of the HS equations, though
having the form of the sum of $Z$-dependent terms, the \rhs of \eq{UpsfullwCCC}
must be  $Z, d Z$-independent. Hence in the vertex analysis  we discard the $ d Z$-dependent terms which are weakly zero anyway.

In this section we present  the final form of the
MNL vertices $\Upsilon^{\eta\bar\eta}_{\go CCC}$ and $\Upsilon^{\eta\bar\eta}_{C\go CC}$.
Technical details are elaborated in  Appendices
  {\bf D} and {\bf E}, respectively.
 The vertices $\Upsilon^{\eta\bar\eta}_{CC\go C}$ and $\Upsilon^{\eta\bar\eta}_{CCC\go }$  can be worked out analogously. (Note that these can be obtained from
the vertices $\Upsilon^{\eta\bar\eta}_{\go CCC}$ and $\Upsilon^{\eta\bar\eta}_{C\go CC}$
by the HS algebra antiautomorphism \cite{Vasiliev:1988sa,Konstein:1989ij,Prokushkin:1998bq}.)

The sketch of the calculation scheme is as follows.

Firstly,  we
write down the  \rhs of     equation \eq{UpsfullwCCC} for $\Upsilon^{\eta\bar\eta}(\go,C,C,C)$.  To this end we use the previously known  $W^\eta _1$, $W^{\bar\eta} _1$,
$B^\eta _2$ and $B^{\bar\eta} _2$
rewritten in the form  \eq{fetabeta} in   Appendix {\bf A},
 MNL $B^{\eta\bar\eta}_3$  of Section~\ref{MNLB3}\,,
$W^{\eta\bar\eta}_2$ obtained in    Appendix {\bf B} in such a way that $[W^{\eta\bar\eta}_2\,,C]_*$ is MNL, and the spin-local vertices  $\Upsilon^{ \eta}(\go,C,C )$   written   in the form \eq{feta} with $\gr=\gb=0$  in Appendix {\bf C}, and their conjugated.

Plugging these terms into the \rhs of    \eq{UpsfullwCCC} one can make sure that the resulting expressions
 have   the form of   Ansatz \eq{fetabeta} and are MNL  for every  ordering of $\go $ and $C$'s.

Let us emphasize that   the full   expression  for  $\Upsilon^{\eta\bar \eta}(\go, C,C,C ) $
\eq{UpsfullwCCC} must be $Z$-independent for each ordering. In principle, one   could  find  manifestly
$Z$-independent expression   by setting for instance
   $Z=0$.
  The result would not be manifestly MNL, since $\gt $
    and $\bgt$   would not be  zero. According to $Z$-dominance Lemma, the $Z$
     dependence can be eliminated by adding
    to the integrand $\dr$-exact expressions giving zero upon integration in the sector in question so that $\gt=\bgt=0$ in the end.
For this we will again use  IH of Section \ref{SimlpeHom}.

Namely, for each ordering, picking up an appropriate  pair  of terms from the \rhs of \eq{UpsfullwCCC} we apply IH dropping  the corresponding $\dr$-exact part. For the
leftover terms, that are  MNL,  this procedure is repeated as many times as needed
until   all of them cancel except for some
 cohomological ones producing the physical vertices.

The resulting MNL vertices  are presented in the next subsections.
Note that it may not be manifest that they  are indeed MNL.
The easiest way to see this is to prove inequalities
\eq{nerav} at the first step of calculations then using repeatedly the simple
inequality
$$|\ga A+(1-\ga)B|+ |\ga A'+(1-\ga)B'|\le
\ga(| A|+|A'|)+ (1-\ga)(|  B|+|B'|)\q \ga\in[0,1]\,.
$$
\subsection{  $\Upsilon^{\eta\bar\eta}_{\go CCC} $ }\label{UpswCCC}
\label{wCCC}
According to \eq{UpsfullwCCC}
\bee \label{UwCCC}
\Upsilon^{\eta\bar\eta}|_{\go CCC}= &&
-(W^{\eta\bar \eta}_2|_{\go CC})*C
-( {W}^{\bar \eta}_1|_{\go C})* B^{  \eta}_2-({W}^{  \eta}_1 |_{\go C})*   B^{\bar \eta}_2
\\ \nn&&- \go* B^{\eta \bar \eta}_3{} -\dr_x B^{\eta \bar \eta}_3{}|_{\go CCC}
-\dr_x B^{  \eta}_2  |_{\go CCC}-\dr_x B^{\bar \eta}_2  |_{\go CCC} \,.
\eee
Using  IH   and formulae
 \eq{B3bulk}, \eq{B3bound},
 \eq{S1withw},  \eq{W1goC}, \eq{B2f}, \eq{bB2f},
 \eq{W2goCCbound}, \eq{UpsgoCC}
    one obtains from Eq.~\eq{UwCCC} moderately non-local
 $\Upsilon^{\eta\bar\eta}|_{\go CCC}$,
  \bee  \label{Upsilon reswCCC}
 \Upsilon^{\eta\bar\eta}|_{\go CCC}= && \ls -\ff{ \eta\bar{\eta}}{16}\!\ls
 \int\limits_{\gt \bgt\gn(2) \ga(2) \gx(3)}
  \!\!   \!\ls\tr(\gt)\tr(\bgt) \mu_1
 \Big[ \EEE(\wmv_1\, |\bwmv_1)
 +  \EEE(\wmv_2\, |\bwmv_2)\Big]\go CCC k\bar{k}
   \\
    \nn
  && \ls+\ff{ \eta\bar{\eta}}{16}\ls \int\limits_{\gt \bgt \gb(2) \ga(2)\gn(2) \gs(2)}
\ls  \!  \!\!\tr(\gt)\tr(\bgt) \mu_2
       \Big[ \EEE(\wmv_3\, |\bwmv_3)
 + \EEE(\wmv_4\, |\bwmv_4)\Big]\go CCC k\bar{k}\quad
  \eee
 where
   $\mu_1 =  \nabla(\ga(2))\nabla(\gn(2)) \nabla(\gx(3))$,
$\mu_2=\nabla(\gb(2)) \nabla(\gn(2))  \nabla(\gs(2)) \nabla(\ga(2))
 $ with $\nabla$ \eq{nabla},
  and
  \bee\label{31w3B3rCH==}
  &&\wmv_1{}^\ga:=- \left [ -(\gn_2+\gn_1\{\ga_2  \gx_2 ({1-\gx_1})^{-1}+\gx_3\}) p_0
  -  \{   \ga_2  \gx_2 ({1-\gx_1})^{-1}+\gx_3\} (p _1{}+ p_2) \right.\qquad\\ \nn&&
 +  \{{\ga_1 \gx_2 }({1-\gx_3})^{-1}  +\gx_1   \}(p_3{}+ p_2) ]^\ga  \q
 \\ \nn
&&\bwmv_1{}^\pa:= -[-(1-\gn_1 \gx_3 ) \bp_0  -(1-\gx_3) (\bp_1{}+\bp_2) + (1-\gx_1)(\bp_3{}+\bp_2) ]^\pa\,,
\\  \label{31w4B3rCH=}
  && \wmv_2{}^\ga:=-[  -(1-\gn_1 \gx_3 )   p_0  - ( 1-\gx_3)  (p _1{}+ p_2) + (1-\gx_1)(p_3{}+ p_2) ]^\ga  \q
 \\ \nn
&&\bwmv_2{}^\pa:=-[ -(\gn_2+\gn_1\{  \gx_3+\ga_1  \gx_2  (1-\gx_1)^{-1} \} ) \bp_0 -
 \{  \gx_3+\ga_1  \gx_2  (1-\gx_1)^{-1} \} (\bp_1{}+\bp_2)
 \qquad\\ &&\nn +\left\{\ga_2 \gx_2(1-\gx_3)^{-1} + \gx_1\right\}(\bp_3{}+\bp_2)]^\pa\,,
  \\\label{31w3B3COH}
  &&\wmv_3{}^\ga:=-
\gb_1   [  -(1-\gn_2 \gs_2\ga_2 )p_0  - ( 1-\gn_2\ga_2)  p _1{}
     +   (-\gn_1 + \gn_2\ga_2)    p_2{}   - \gn_1    p_3{}]^\ga -p_3{}^\ga \q
 \\ \nn
&&\bwmv_3{}^\pa:=
-[-(1-\gs_2 \ga_1) \bp_0  - \ga_2 (\bp_1{}+\bp_2) +  (\bp_3{}+\bp_2) ]^\pa\,,
  \\\label{31w4B3rCOH}
  &&
 \wmv_4{}^\ga:=-[-(1-\gs_2 \ga_1)  p_0  - \ga_2 ( p_1{}+ p_2) +  ( p_3{}+ p_2) ]^\ga   \q
 \\ \nn
&&\bwmv_4{}^\pa:= -\gb_1   [
  -(1-\gn_2 \gs_2\ga_2 )\bp_0  - ( 1-\gn_2\ga_2)  \bp _1{}
    +   (  \gn_2(\ga_2+1)- 1)    \bp_2{}- \gn_1   \bp_3{} ]^\pa   - \bp_3{}^\pa.
 \qquad\eee

\subsection{ $\Upsilon^{\eta\bar\eta}_{  C\go CC} $}
\label{CwCC}  \label{fcohomoloCwCCC}
  According to \eq{UpsfullwCCC},
\bee \label{UCwCC}
\Upsilon^{\eta\bar\eta}|_{C\go CC}= &&
C*(W^{\eta\bar \eta}_2|_{\go CC})
-( W^{\eta\bar \eta}_2|_{C\go C})*C -( {W}^{\bar \eta}_1* B^{  \eta}_2
+{W}^{  \eta}_1*   B^{\bar \eta}_2)|_{C\go  CC}
\\ \nn&&-\dr_x B^{\eta \bar \eta}_3{}|_{C \go CC}
-\dr_x B^{  \eta}_2  |_{C\go CC}-\dr_x B^{\bar \eta}_2  |_{C\go CC} \,.
\eee

 Using  IH and formulae
\eq{B3bulk}, \eq{B3bound},
 \eq{S1withw}, \eq{W1goC}, \eq{W1Cgo},  \eq{B2f}, \eq{bB2f},
  \eq{W2goCCbound},
 \eq{W2CgoCbound} and \eq{UpsgoCC}, \eq{UpsCgoC}
one obtains from  Eq.~\eq{UCwCC}  moderately non-local
 $\Upsilon_{C\go CC}$ \,,
\bee   &&\Upsilon^{\eta\bar\eta\,\,}_{  C\go CC} =
\ff{\eta\bar{\eta}}{16}\!\! \int\limits_{\gt \bgt\gn(2) \ga(2) \gx(3)}
  \!\!   \!\!\tr(\gt)\tr(\bgt)   \label{ResultCgoCCb1}      \mu_1\Big[\EEE(\wmv_1\, |\bwmv_1)
+\EEE(\wmv_2\, |\bwmv_2)\Big]
C\go CC k\bar{k}  \\ \nn
&&
 +\ff{ \eta\bar{\eta}}{16}\!\!   \!\!\!\!\!\!
   \!\! \int\limits_{\gt \bgt \gb(2) \ga(2)\gn(2) \gs(2)}
    \!\!\!\!\!\!\tr(\gt)\tr(\bgt)  \mu_2\Big[\EEE(\wmv_3\, |\bwmv_3)
+\EEE(\wmv_4\, |\bwmv_4)-\EEE(\wmv_5\, |\bwmv_5)
 -\EEE(\wmv_6\, |\bwmv_6)
  \Big]
C\go CC k\bar{k}\,  \eee
with $\mu_1 =  \nabla(\ga(2))\nabla(\gn(2)) \nabla(\gx(3))$,
$\mu_2=\nabla(\gb(2)) \nabla(\gn(2))  \nabla(\gs(2)) \nabla(\ga(2))
     \,  $ and
\bee\label{CgoCCw3B3r=}
 &&\wmv_1{}^\ga : =-\left [  (\{{\ga_2 \gx_2 }({1-\gx_3})^{-1}
  +\gx_1   \}- \gn_1\{\ga_1  \gx_2 ({1-\gx_1})^{-1}+\gx_3\}) p_0
 \right. \\ \nn&&
 \!\left.-  \{   \ga_1  \gx_2 ({1-\gx_1})^{-1}+\gx_3\} (p _1{}+ p_2)
 +  \{{\ga_2 \gx_2 }({1-\gx_3})^{-1}  +\gx_1   \}(p_3{}+ p_2) \right]^\ga  \q
  \\ \nn
&&\bwmv_1{}^\pa:= -[ (\gn_1(1-\gx_1)- (1-\gx_3)) \bp_0  -(1-\gx_3) (\bp_1{}+\bp_2)
+ (1-\gx_1)(\bp_3{}+\bp_2) ]^\pa\,,
 \\ \label{31w4B3r=}
 &&
 \wmv_2{}^\ga:=-[  (\gn_1(1-\gx_1)-(1-\gx_3)) p_0  - ( 1-\gx_3)  (p _1{}+ p_2) + (1-\gx_1)(p_3{}+ p_2) ]^\ga  \q
 \\ \nn
&&\bwmv_2{}^\pa := -[  (\gn_1\{\ga_1 \gx_2(1-\gx_3)^{-1} + \gx_1 \}
-\{  \gx_3+\ga_2  \gx_2  (1-\gx_1)^{-1} \} ) \bp_0
 \\ &&\nn- \{  \gx_3+\ga_2  \gx_2  (1-\gx_1)^{-1} \} (\bp_1{}+\bp_2)
 + \{\ga_1 \gx_2(1-\gx_3)^{-1} + \gx_1 \}(\bp_3{}+\bp_2) ]^\pa\,,
   \\ \label{Cohgt03=}  &&
  \wmv_3{}^\ga := \!-\!
( [ \gn_2 \!-\!\gs_1  ] p_0\!-\!\gs_1 p_1+\gs_2 p_2 + p_3)^\ga
 \q\\ \nn
&&\bwmv_3 {} ^\pa := \!-\!
  \{  \gb_1(  \ga_1[ \gs_1    \!-\!\gn_1   ] \bp_0\!-\!\ga_1\gs_2\bp_1+\ga_1\gs_1\bp_2 +
 \bp_3)+ \gb_2(\!-\!\ga_1\bp_0    \!-\!\ga_1  \bp _1\!-\!\ga_1 \bp_2
+ \ga_2\bp_3)
  \}^\pa \, ,\qquad \\ \label{Cohgt03b}  &&
 \wmv_4{} ^{ \gn} := \!-\!
 \{  \gb_1(  \ga_1[ \gs_1    \!-\!\gn_1   ] p_0\!-\!\ga_1\gs_2p_1+\ga_1\gs_1p_2 +
 p_3)+
  \gb_2(\!-\!\ga_1p_0    \!-\!\ga_1  p _1\!-\!\ga_1 p_2
+ \ga_2p_3)
  \}^\ga ,\qquad \\ \nn
&& \bwmv_4 {}^{ \pn} := \!-\!
( [ \gn_2 \!-\!\gs_1  ] \bp_0\!-\!\gs_1 \bp_1+\gs_2 \bp_2 + \bp_3)^\pa ,\qquad
 \\\label{Cohgt01=}
&&\wmv_5{}^{\ga}  :=
\!-\![  (\gn_2\gs_2 \!-\! 1 ) p_0\!-\!p_1\!-\!\gn_1  p _2 + \gn_2 p_3   ]{}^\ga   ,\qquad  \\ \nn
&& \bwmv_5{}^\pa{}  :=  \!-\![
( \ga_{1 } +\gb_2 (\!-\!\gs_1\!-\!\gs_2 \ga_2 \gn_2
     ))\bp_0\!-\! \ga_{2 }  \bp _1
  +   \{\!-\! \gb_2 \ga_2  \gn_2   +\ga_1 \}   \bp_2
 + \{ \gb_2\ga_2 \gn_1   +\ga_1\} \bp_3{} ]^\pa
  \,,\\\label{Cohgt02b}
&& \wmv_6{}^\ga{}  :=  \!-\![
( \ga_{1 } +\gb_2 (\!-\!\gs_1\!-\!\gs_2 \ga_2 \gn_2
     )) p_0   \!-\!  \ga_2     p_1{}
       + \{\!-\! \gb_2 \ga_2  \gn_2   +\ga_1 \}   p_2{}+\{ \gb_2\ga_2 \gn_1   +\ga_1\}p_3]^\ga
 \q\,\,\,\\ \nn
 &&\bwmv_6{}^\pa  :=
\!-\![  ( \gn_2\gs_2 \!-\! 1 ) \bp_0\!-\!\bp_1\!-\!\gn_1  \bp _2 + \gn_2 \bp_3   ]{}^\pa
  \,.\eee

   \subsection{ $\Upsilon^{\eta\bar\eta}_{  CC\go C} $}
\label{CCwC}
     \label{fcohomoloCCwC}
According to  \eq{UpsfullwCCC},
\bee \label{UCCwC}
\ls\Upsilon^{\eta\bar\eta}|_{CC\go C}= &&
C*(W^{\eta\bar \eta}_2|_{C\go C})
- (W^{\eta\bar \eta}_2|_{CC\go })*C +(   B^{  \eta}_2*{W}^{\bar \eta}_1+ B^{\bar \eta}_2*{W}^{  \eta}_1)|_{CC\go  C}
\\ \nn&&-\dr_x B^{\eta \bar \eta}_3{}|_{CC\go  C}
-\dr_x B^{  \eta}_2  |_{CC\go  C}-\dr_x B^{\bar \eta}_2  |_{CC\go  C} \,.
\eee
Using  IH   and formulae \eq{B3bulk}, \eq{B3bound},  \eq{S1withw}, \eq{W1goC}, \eq{W1Cgo}, \eq{B2f},
\eq{bB2f},
 \eq{W2CgoCbound},
  \eq{W2CCgobound},
  \eq{UpsCgoC}, \eq{UpsCCgo}
 one obtains from  Eq.~\eq{UCCwC} moderately non-local
 $\Upsilon_{CC\go C}$\,,
\bee   &&\Upsilon^{\eta\bar\eta\,\,}_{CC\go C} =
-\ff{\eta\bar{\eta}}{16}\!\! \int\limits_{\gt \bgt\gn(2) \ga(2) \gx(3)}
  \!\!   \!\!\tr(\gt)\tr(\bgt)
   \label{ResultCCwCb1}
    \mu_1\Big[\EEE(\wmv_1\, |\bwmv_1)
+\EEE(\wmv_2\, |\bwmv_2)\Big] CC\go C k\bar{k}\,
\\ \nn
&&
-\ff{ \eta\bar{\eta}}{16}\!\!   \!\!\!\!\!\!
   \!\! \int\limits_{\gt \bgt \gb(2) \ga(2)\gn(2) \gs(2)}
   \!\!\!\!\!\!\tr(\gt)\tr(\bgt) \mu_2\Big[\EEE(\wmv_3\, |\bwmv_3)
+\EEE(\wmv_4\, |\bwmv_4)-\EEE(\wmv_5\, |\bwmv_5)
  -\EEE(\wmv_6\, |\bwmv_6)
  \Big]  CC\go C k\bar{k}\, \qquad  \eee
with
$\mu_1 =  \nabla(\ga(2))\nabla(\gn(2)) \nabla(\gx(3))$, $
\mu_2=\nabla(\gb(2)) \nabla(\gn(2))  \nabla(\gs(2)) \nabla(\ga(2))
     \,, $
 \bee\label{BCCgoCw3B3rCoh}
 &&\wmv_1{}^\ga :=-\left [  (\{{\ga_2 \gx_2 }({1-\gx_3})^{-1}
  +\gx_1   \}- \gn_1\{\ga_1  \gx_2 ({1-\gx_1})^{-1}+\gx_3\}) p_0
 \right. \\ \nn&&
 \left.-  \{   \ga_1  \gx_2 ({1-\gx_1})^{-1}+\gx_3
   \} (p _1{}+ p_2)
 +  \{{\ga_2 \gx_2 }({1-\gx_3})^{-1}
  +\gx_1   \}(p_3{}+ p_2) \right]^\ga  \q
 \\ \nn
&&\bwmv_1{}^\pa:= -[ ((1-\gx_1)- \gn_1(1-\gx_3)) \bp_0  -(1-\gx_3) (\bp_1{}+\bp_2)
+ (1-\gx_1)(\bp_3{}+\bp_2) ]^\pa\,,
 \\
  \label{B31w4B3rcoh}
 &&
 \wmv_2{}^\ga:=-[  ((1-\gx_1)-\gn_1(1-\gx_3)) p_0
  - ( 1-\gx_3)  (p _1{}+ p_2) + (1-\gx_1)(p_3{}+ p_2) ]^\ga  \q
 \\ \nn
&&\bwmv_2{}^\pa := -[  (\{\ga_1 \gx_2(1-\gx_3)^{-1} + \gx_1 \}
-\gn_1\{  \gx_3+\ga_2  \gx_2  (1-\gx_1)^{-1} \} ) \bp_0
 \\ &&\nn- \{  \gx_3+\ga_2  \gx_2  (1-\gx_1)^{-1} \} (\bp_1{}+\bp_2)
 + \{\ga_1 \gx_2(1-\gx_3)^{-1} + \gx_1 \}(\bp_3{}+\bp_2) ]^\pa\,,
 \eee
 \bee\label{UnifCCwC1CH}
&&\wmv_3{}^\ga{}  :=
\!-\![  (\!-\!\gs_1   + \gn_2 ) p_0\!-\! p_1\!-\!\gs_1  p _2 + \gs_2 p_3   ]{}^\ga  \,, \\ \nn
&& \bwmv_3{}^\pa{}  : =  \!-\!\big[ \ga_2 ( 1
  \!-\!\gb_2( \gs_2+\gn_1  )\bp_0 \!-\!(1\!-\!\ga_2 \gb_1 )\bp_1
  + \ga_2 (\gb_1\!-\!\gs_2 \gb_2) \bp_2+\ga_2(1\!-\!\gs_2\gb_2)\bp_3
 \big]^\pa,
\\\label{UnifCCwC1CHb}
 &&  \wmv_4{}^\ga{}  : =  \!-\!\big[ \ga_2 ( 1
  \!-\!\gb_2( \gs_2+\gn_1  )  p_0\!-\!(1\!-\!\ga_2 \gb_1 ) p _1+ \ga_2 (\gb_1\!-\!\gs_2 \gb_2)  p_2
  +\ga_2(1\!-\!\gs_2\gb_2) p_3 )
 \big]^\ga,
\qquad\\\nn &&\bwmv_4{}_{\pa}{}  :=
\!-\![  (\!-\!\gs_1   + \gn_2 ) \bp_0\!-\! \bp_1\!-\!\gs_1  \bp _2 + \gs_2 \bp_3   ]{}^\pa \,,
   \\
  \label{Cohgt01CCwC}
  &&
 \wmv_5 {}^\ga:=\!-\![  (1\!-\!\gs_1 \gn_1) p_0  \!-\!  \gn_1  (p _1{}+ p_2) +  (p_3{}+ p_2) ]^\ga   ,\qquad  \\ \nn
&&\bwmv_5 {}^\pa :=\!-\![ (  \!-\!\ga_1
+\gb_2\gs_2+ \gb_2\gs_1\ga_2  \gn_1     )\bp_0 \!+\! ( \!-\! \ga_1\!-\! \ga_2  \gb_2 \gn_2  ) \bp _1
 \!+\! \{\!-\! \ga_1 \!+\! \gb_2 \ga_2  \gn_1    \}\bp_2
 \!+\! \ga_2      \bp_3{} ]^\pa ,\qquad
\\ \label{Cohgt01CCwCb}
  &&
 \wmv_6 {}^\ga:=\!-\![ (  \!-\!\ga_1
\!+\!\gb_2\gs_2+ \gb_2\gs_1\ga_2  \gn_1     )p_0 \!+\! ( \!-\! \ga_1\!-\! \ga_2  \gb_2 \gn_2  ) p _1
 \!+\! \{\!-\! \ga_1 \!+\! \gb_2 \ga_2  \gn_1    \}p_2
 \!+\! \ga_2      p_3{} ]^\ga   ,\qquad  \\ \nn
&&\bwmv_6 {}^\pa :=\!-\![  (1\!-\!\gs_1 \gn_1) \bp_0  \!-\!  \gn_1  (\bp _1{}\!+\! \bp_2)
\!+\!  (\bp_3{}+ \bp_2) ]^\pa .
  \eee

    \subsection{ $\Upsilon^{\eta\bar\eta}_{  CCC\go} $}
 \label{CCCw}

According to \eq{UpsfullwCCC},
\bee \label{UCCCw}
\Upsilon^{\eta\bar\eta}|_{CCC \go }= &&
C*(W^{\eta\bar \eta}_2|_{  CC\go })
+   B^{  \eta}_2 *({W}^{\bar \eta}_1|_{C\go})+    B^{\bar \eta}_2*({W}^{  \eta}_1 |_{C\go})
\\ \nn&&+ B^{\eta \bar \eta}_3{}*\go
- \dr_x B^{\eta \bar \eta}_3{}|_{CCC\go}
-\dr_x B^{  \eta}_2  |_{CCC\go}-\dr_x B^{\bar \eta}_2  |_{CCC\go} \,.
\eee
Using  IH   and formulae
\eq{B3bulk}, \eq{B3bound},  \eq{S1withw}, \eq{W1Cgo}, \eq{B2f}, \eq{bB2f},
  \eq{W2CCgobound}, \eq{UpsCCgo}
 one obtains from  Eq.~\eq{UCCCw} moderately non-local
 $\Upsilon_{CCC\go}$
     \bee
\Upsilon_{  CCC\go} =\label{COHOMDB3CCCw=} &&
\ff{\eta\bar{\eta}}{16}\!\! \int\limits_{\gt \bgt\gn(2) \ga(2) \gx(3)}
 \!\!   \!\!\tr(\gt)\tr(\bgt)  \mu_1
 \Big[\EEE(\wmv_1\, |\bwmv_1)
+\EEE(\wmv_2\, |\bwmv_2)\Big]   CCC\go k\bar{k}
  \\ \nn
  &&-\ff{ \eta\bar{\eta}}{16}\!\!   \!\!\!\!\!\!   \!\! \int\limits_{\gt \bgt \gb(2) \ga(2)\gn(2) \gs(2)}
\!\!   \!\!\tr(\gt)\tr(\bgt) \mu_2
\Big[ \EEE(\wmv_3\, |\bwmv_3)
 + \EEE(\wmv_4\, |\bwmv_4)\Big] CCC \go k\bar{k}\q
  \eee
with $\mu_1 =  \nabla(\ga(2))\nabla(\gn(2)) \nabla(\gx(3))$,
$
\mu_2=\nabla(\gb(2)) \nabla(\gn(2))  \nabla(\gs(2)) \nabla(\ga(2))
     \,  $
and
\bee\label{31w3B3rCCCgCOH}
  &&\wmv_1^{ \ga}:= -
 \left [ (\gn_1+\gn_2\{{\ga_2 \gx_2 }({1-\gx_3})^{-1}+\gx_1  \}) p_0\right.
 \\ \nn&&
\!\quad\qquad \left.  -  \{   \ga_1  \gx_2 ({1-\gx_1})^{-1}+\gx_3
   \} (p _1{}+ p_2)^\ga+
   \{{\ga_2 \gx_2 }({1-\gx_3})^{-1}+\gx_1   \}(p_3{}+ p_2)^\ga\right]  \q
 \\ \nn
&&\bwmv_1^{ \pa}:= -[(\gn_1+\gn_2(1-\gx_1)) \bp_0  -(1-\gx_3) (\bp_1{}+\bp_2) + (1-\gx_1)(\bp_3{}+\bp_2) ]^\pa\,,
\\  \label{31w4B3rCCCgCOH}
  &&
 \wmv_2^{ \ga}:=-[  (\gn_1+\gn_2(1-\gx_1)) p_0  - ( 1-\gx_3)  (p _1{}+ p_2)+ (1-\gx_1)(p_3{}+ p_2) ]^\ga  \q
 \\ \nn
&&\bwmv_2^{ \pa}:= -
[ (\gn_1+\gn_2\{ \ga_2 \gx_2(1-\gx_3)^{-1} + \gx_1 \} ) \bp_0
\\ &&\nn
\!\quad\qquad-
 \{   \ga_1  \gx_2  (1-\gx_1)^{-1} +\gx_3\} (\bp_1{}+\bp_2)
  +\left\{\ga_2 \gx_2(1-\gx_3)^{-1} + \gx_1\right\}(\bp_3{}+\bp_2) ]^\pa
\,,
  \\ \label{lastw3CCCWCOH}  && \wmv_3{}^\ga:= -[ ( 1-\gs_1\gn_1)p_0-p_1  -\gn_1  p _2{}+    \gn_2    p_3{}
 ]^\ga  \q
\\   \nn
&&\bwmv_3{}^\pa :=-\big(  \gb_2 (1-\ga_1 \gs_1\gn_2 )\bp_0
-(1-\ga_2\gb_2)\bp_1
+\gb_2(\ga_2- \ga_1\gn_2 )\bp_2{}+ \gb_2(1-\ga_1 \gn_2 )\bp_3{}\big)^\pa\,\q
\\ \label{lastw4CCCWCOH}
  &&\wmv_4{}^\ga := -\big(
\gb_2 (1-\gs_1\gn_2\ga_1)p_0-(1-\ga_2\gb_2)p_1
+\gb_2(\ga_2- \ga_1\gn_2)   p_2{}+ \gb_2(1-\ga_1 \gn_2 )    p_3{}
  \big) ^\ga \q
\\ \nn &&\bwmv_4{}^\pa := -[  ( 1-\gs_1\gn_1)\bp_0-\bp_1 - \gn_2 \bp_2{}+ \gn_1 \bp_3{}
 ]^\pa\,.
 \eee

\section{Conclusion}
\label{Con}
In this paper we introduce the concept
of moderate non-locality and calculate moderately non-local
vertices  $\Upsilon^{\eta\bar \eta}(\go, C,C,C)$ in the mixed $\eta\bar\eta$ sector
of HS gauge theory in $AdS_4$ for all orderings of the fields  $\go$ and $C$.
Our calculation is based on the  differential  homotopy   Ansatz of \cite{Vasiliev:2023yzx} for the  lowest order holomorphic deformation linear in $\eta$
of the perturbative analysis of the holomorphic sector. To solve the problem
we use the interpolating homotopy  that preserves  moderate non-locality in the process of perturbative analysis of the HS equations.

The  degree of non-locality  of  vertices is expressed by the
coefficients $ P^{ij}$ and $ \bar P^{ij}$   in front of, respectively,
convolutions $p_i{}_\ga p_j{}^\ga $ and $\bp_i{}_\pa \bp_j{}^\pa$ in the exponents $\EEE$    \eq{EEE}. Moderately non-local vertices obey the inequalities $|P^{ij}|+ |\bar P^{ij}| \le  1$,
while the usual star product $C_1(y,\by)*C_2(y,\by)*  C_3(y,\by)$ yields $|P^{ij}|+|\bar P^{ij}|=2$.
At the moment, moderately non-local vertices are minimally non-local among  known
vertices in the mixed $\eta\bar\eta$ sector of the HS gauge theory. Note that the usual spin-local vertices of \cite{Vasiliev:2016xui} in the (anti)holomorphic sector
 form a subclass  of moderately non-local vertices. Let us also stress that our
 construction is manifestly invariant under HS gauge symmetries.

The results of this paper raise a number of interesting questions for the future study.
The most important one is to understand whether it is possible to improve further
the level of non-locality of HS theory by choosing appropriate field variables. Another interesting problem is to compare the
level of non-locality of the moderately non-local vertices  with that deduced
by Sleight and Taronna \cite{Sleight:2017pcz} from the
Klebanov-Polyakov holographic conjecture \cite{Klebanov:2002ja}.

It is also important to extend the results of this paper to the vertex
$\Upsilon^{\eta\bar\eta}(\go,\go,C,C)$.
 Presumably, spin-local $S_2^{\eta\bar \eta}$
 and $W_2^{\eta\bar \eta}$ obtained in this paper    lead  to the special
 form of the local  bilinear $ {\eta\bar \eta}-$current deformation  in the
one-form sector, originally obtained in
\cite{Gelfond:2017wrh} using conventional homotopy
supplemented by some field redefinitions, that  leads  to
the current contribution to Fronsdal equations \cite{Misuna:2017bjb} in agreement
with  Metsaev's classification
 \cite{Metsaev:1991mt,Metsaev:2007rn}.

Moreover,   the IH  approach of this paper makes it possible to
obtain the spin-local  vertex $\Upsilon^{\eta\bar\eta}(\go,\go,C,C)$ such that  $[\Upsilon^{\eta\bar\eta}(\go,\go,C,C),C]_*$ is MNL.
(Note that  $[\tilde{\Upsilon}^{\eta\bar\eta}(\go,\go,C,C),C]_*$ is not MNL for  the spin-local vertex
$ \tilde{\Upsilon}^{\eta\bar\eta}(\go,\go,C,C) $ obtained in \cite{Didenko:2019xzz}.)
This property is important for the analysis of the   contribution  of the vertices $\Upsilon^{\eta\bar\eta} $ to Fronsdal equations.

The sketch of the calculation   is as follows.
Eq.~\eqref{HS3} yields
\begin{equation}
\label{UpcgogoCC}
\Upsilon^{\eta\bar\eta}(\go,\go,C,C)=
-\big(\mathrm{d}_x
W_1+W_1* W_1+\mathrm{d}_x W_2+\omega* W_2+W_2*
\omega\big)\big|_{{\eta\bar\eta}}.
\end{equation}
Plugging  $W^\eta _1$,  $W^{\bar\eta} _1$ (\eq {W1goC}, \eq{W1Cgo}), $W^{\eta\bar\eta} _2$
(\eq {W2goCCbound}, \eq{W2CgoCbound}\,, \eq{W2CCgobound})  into the \rhs of \eq{UpcgogoCC}  along with Eqs.~\eq{UpsgoCC}-\eq{UpsCCgo} and their conjugated one can make sure
 that $\Upsilon^{\eta\bar\eta}(\go,\go,C,C)$ \eq{UpcgogoCC} is spin-local and
 $[\Upsilon^{\eta\bar\eta}(\go,\go,C,C),C]_*$ is MNL for every  ordering of $\go $ and $C$'s.
Hence, to eliminate $Z$-dependence in a way preserving MNL, one  can again
use  IH of Section \ref{SimlpeHom}.
This is  work in progress.

 \section*{Acknowlegement}
I am  grateful to Mikhail Vasiliev for careful reading the manuscript and  many
useful discussions and comments.
I wish to thank for hospitality  Ofer Aharony, Theoretical High Energy Physics Group
of Weizmann Institute of Science where some part of this work was done.

 \newcounter{appendix}
\setcounter{appendix}{1}
\renewcommand{\theequation}{\Alph{appendix}.\arabic{equation}}
\addtocounter{section}{1} \setcounter{equation}{0}
 \renewcommand{\thesection}{\Alph{appendix}.}
 \addcontentsline{toc}{section}{\,\,\,\,\,\,\,Appendix A:  $S^{\eta(\bar\eta)}_1$ ,$W^{\eta(\bar\eta)}_1$, $B^{\eta(\bar\eta)}_2$  }

\section*{Appendix A:   $S _1$, $W _1$, $B _2$}
% and mixed $W^{\eta\bar\eta} _2$     }
\label{sectlowSWB}

  We rewrite  $S_1$ ,$W_1$ and   $B_2$   obtained in
\cite{Didenko:2018fgx, Didenko:2019xzz}  via  $\EE(...)$ defined in
\eq{Exp} and its conjugated
$\bEE(...)$  (see also \cite{Vasiliev:2023yzx})
 \bee\label{S1withw}
&&  S^\eta_1\, =
 -\ff{\eta}{2 }\int_{\gt  } l(\gt)
  \EE(\wmv ) C * k-\ff{\bar{\eta}}{2 }\int_{\bgt  } l(\bgt)   \bEE(\bwmv) C * \bk\q
 \\\label{wS1withw}
&& \wmv^\ga= \gt z^\ga \q
 \bwmv^\ga=\bgt \bz^\pa \,,\eee
 \bee\label{W1goC}
\ls W_1 \,|_{\go C} &=& \ff{ i\eta}{4 }
\!\!\int\limits_{ \gt,   \gs  } l(\gs)    l(\gt)
 \EE(\wmv)  \go  C  *k
+
\ff{i\bar\eta}{4 }\!\!\int\limits_{  \gt,  \gs  } l (\gs)
 l(\bgt)     \bEE(\bwmv)  \go  C   *\bk
\q  \\
  \label{wW1goC}&&\wmv^\ga =\gt z^\ga-(1-\gt) (- \gs   p_0 )^\ga  \q
 \bwmv^\pa =\bgt \bz^\pa-(1-\bgt)(- \gs  \bp_0 )^\pa
 \,\q\eee
\bee\label{W1Cgo}
  W_1\,|_{C\go } &= &\ff{\eta}{4i }
\!\int\limits_{   \gt  \gs  } l (\gs) l(\gt)     \EE(\wmv)     C   \go *k
+
 \ff{ \bar\eta}{4i }
   \int\limits_{  \gt,  \gs } l(\gs) l(\bgt) \bEE(\bwmv) C  \go    *\bk
\q  \\
 \label{wW1Cgo}&&\wmv^\ga =
\gt z^\ga-(1-\gt) (\gs   p_0 )^\ga
  \q  \bwmv^\pa =\bgt \bz^\pa-(1-\bgt)
 (\gs \bp_0 )^\pa
 \q\eee
     \bee \label{B2f}
B_{2}^{\eta}  &=&
\!\!\ff{\eta}{4i} \int\limits_{    \gt   \gs(2) } l(\tau)  \nabla(\gs(2))
   \EE(\wmv)\,C  C * k\,,
\\  \label{wrtB2}
  \wmv^\ga  &=&\tau z^\ga -(1- \tau) (\gs_2 p_2 -\gs_1 p _1)^\ga \,,
\eee
 \bee\label{bB2f}
B_{2}^{\bar\eta} &=&
\!\! \ff{\bar\eta}{4i}\int\limits_{    \bgt  \gs(2) } l(\bgt) \nabla(\gs(2))
   \bEE(\bwmv)\,\,C  C  *\bar k\,,
\\\label{wrtB2b}
 \bwmv^\pa  &=&
  \bgt \bz^\pa -(1- \bgt) (\gs_2 \bp_2 -\gs_1 \bp _1)^\pa \,.
\eee

 \addtocounter{appendix}{1}
\renewcommand{\theequation}{\Alph{appendix}.\arabic{equation}}
\addtocounter{section}{1} \setcounter{equation}{0}
 \addcontentsline{toc}{section}{\,\,\,\,\,\,\,Appendix B:   $W_2^{\eta\bar\eta}$ }
 \section*{Appendix B:   $W_2^{\eta\bar\eta}$}
\label{w2MNL}
Here we construct an appropriate $W_2^{\eta\bar\eta}$ using
  $S_2^{\eta\bar\eta}$ \eq{10intsS2}.

 One can make sure straightforwardly that if $\dr {W}^{\eta \bar\eta}_2$
satisfies \eq{dW2=} with $S_2$  \eq{10intsS2}
then
$[\dr {W}^{\eta \bar\eta}_2, C]_*$  is MNL by virtue of
    Eqs~\eq {S1withw}-\eq{bB2f} and \eq{10intsS2}.
Hence, using the technique  of  \cite{Vasiliev:2023yzx} and IH we obtain   such  $W_2^{\eta\bar\eta}$
that  $[W_2^{\eta\bar\eta}\,,C]_*$ is MNL,
 \bee\nn\ls &&
 W^{\eta\bar{\eta}}_2 {}_{\go C{C} }{}  =     \ff{\eta\bar{\eta}}{16}
\int\limits_{\gt\,, \bgt\,, \gs(2)  \, \gn(2)  }
 \nabla(\gs(2))   \nabla(\gn(2))
\times \nn
\Bigg[\Box (\gt,\bgt)  \EEE (\wmv_1  | \bar  \wmv_1 )
\\ \label{W2goCCbound}&&
\ls+\int\limits_{  \ga(2)} \nabla(\ga(2))   \Big\{
-\tr (\gt)l(\bgt) \EEE (\wmv_2  | \bar  \wmv_2 )
-\tr (\bgt)l(\gt)  \EEE (\wmv_3  | \bar  \wmv_3 )
\Big\}\Bigg]\go CC k\bar{k} \,,\quad
\eee
 where
\bee\label{w1goCCbu}  &&
 \wmv_1{}^\ga:= \gt  z^\ga- (1-\gt )[ -(\gn_1+\gn_2\gs_1)p_0   -\gs_1  p_1{} +    \gs_2    p_2{}
]^\ga  \q
\\ \nn
&&\bar \wmv_1{}^\pa:= \bgt  \bz^\pa -(1-\bgt )[
-(\gn_1+\gn_2\gs_2)\bp_0 - \gs_2 \bp_1{} + \gs_1 \bp_2{}  ]^\pa\,,
 \\   \label{W2gtgt0=}
&& {\wmv_2 }{}^\ga:= -[-(\gn_1+\gn_2\gs_1)p_{0}  -\gs_1  p _1{}+    \gs_2    p_2{}
 ]^\ga  \q
\\ \nn
&&\bwmv_{2}{}^\pa:= \bgt  \bz^\pa -(1-\bgt )\ga_1[-(\gn_1+\gn_2\gs_2)\bp_{0}- \gs_2 \bp_1{} + \gs_1 \bp_2{}
 ]^\pa\,\q
 \\\label{w3goCCbgt0=}&&\wmv_3{}^\ga:= \gt  z^\ga- (1-\gt )\ga_1[
-(\gn_1+\gn_2\gs_1)p_{0}   - \gs_1  p_1{}+     \gs_2    p_2{}
 ] ^\ga \q
\\ \nn
&&\bwmv_3{}^\pa:= -[- (\gn_1+\gn_2\gs_2)\bp_{0} - \gs_2 \bp_1{}+ \gs_1 \bp_2{}
 ]^\pa\,,
\eee
\bee
 \ls&&
 W^{\eta\bar{\eta}}_2 {}_{C\go C} =   \ff{\eta\bar{\eta}}{16}
\ls \int\limits_{\gt\,  \bgt  \gn(2)  \gs(2)  }
\ls    \nabla(\gn(2))  \nabla(\gs(2))
\times \nn
\Bigg [\Box (\gt,\bgt)
\big( \EEE (\wmv_0 | \bar  \wmv_0  )+\EEE (\wmv_1  | \bar  \wmv_1 )
 +\EEE (\wmv_2  | \bar  \wmv_2 )\big)  \,
\\ \label{W2CgoCbound}&&  \ls+
\int\limits    _{  \ga(2)} \nabla(\ga(2))
    \Big\{
\tr (\gt)l(\bgt)  \EEE (\wmv_3  | \bar  \wmv_3 )
-\tr (\bgt)l(\gt)  \EEE (\wmv_4  | \bar  \wmv_4 )
\Big\}\Bigg]C\go C k\bar{k} \,,
\eee
where
\bee \label{OW2CgoC0}  &&\wmv_0{}^\ga = \gt  z^\ga- (1-\gt )[ (-\gs_1 \gn_1+\gs_2 \gn_2)p_0 -\gs_1  p _1{} +    \gs_2    p_2{}  ]^\ga   \q
\\ \nn
&&\bwmv_0{}^\pa = \bgt  \bz^\pa -(1-\bgt )[(-\gs_2 \gn_1+\gs_1 \gn_2)  \bp_{ 0}- \gs_2 \bp_1{}
+ \gs_1 \bp_2{}   ]^\pa
\q\eee
\bee \label{OW2CgoC1} &&\wmv_1{}^\ga = \gt  z^\ga- (1-\gt )[ -(  \gs_1 \gn_2+  \gn_1) p_0 -   p _1{}   ]^\ga   \q
\\ \nn
&&\bwmv_1{}^\pa= \bgt  \bz_1{}^\pa -(1-\bgt )[( \gs_1 \gn_1+  \gn_2) \bp_{ 0}+\bp_2{}   ]^\pa\, \q
\\
 \label{OW2CgoC2}
 &&\wmv_2{}^\ga= \gt  z^\ga- (1-\gt )[ (  \gs_2 \gn_1+  \gn_2) p_0 +   p _2{}   ]^\ga   \q
\\ \nn
&&\bwmv_2{}^\pa= \bgt  \bz^\pa -(1-\bgt )[-( \gs_2 \gn_2+  \gn_1) \bp_{ 0}-\bp_1{}   ]^\pa\,\q
\\
 \label{PW2CgoC131==}  && \wmv_{3}{}^\ga= -
( [\gn_2\gs_2 -\gn_1 \gs_1 ] p_0-\gs_1 p_1+\gs_2 p_2 )^\ga
\q\\ \nn
&&\bwmv_{3}{} ^\pa= \bgt  \bz^\pa- (1-\bgt )
\ga_1   ([\gn_2\gs_1    -\gn_1\gs_2  ] \bp_0-\gs_2\bp_1+\gs_1\bp_2 )^\pa \q
\\
 \label{PW2CgoC131=4}
&& \wmv_{4}{} ^\ga=  \gt  z^\ga- (1-\gt )
\ga_1   ([\gn_2\gs_1    -\gn_1\gs_2  ]  p_0-\gs_2 p_1+\gs_1 p_2
 )^\ga\q\\ &&\nn
  \bwmv_{4}{}^\pa= -
( [\gn_2\gs_2 -\gn_1 \gs_1 ] \bp_0-\gs_1 \bp_1+\gs_2 \bp_2 )^\pa
  \q
 \eee

\bee\nn&&
 W^{\eta\bar{\eta}}_2 {}_{CC\go   }{}  =    \ff{\eta\bar{\eta}}{16}
\int\limits_{\gt\,  \bgt\,  \gs(2)   \gn(2)  }
\ls\nabla(\gs(2))\nabla(\gn(2))
\times\Bigg[
 \Box (\gt,\bgt) \EEE (\wmv_1  | \bar  \wmv_1 )
   \,
\\\label{W2CCgobound}&& \ls \!-\!
\int\limits_{  \ga(2)} \nabla(\ga(2))  \Big\{
\tr (\gt)l(\bgt) \EEE (\wmv_2  | \bar  \wmv_2 )
\!+\!\tr (\bgt)l(\gt)  \EEE (\wmv_3  | \bar  \wmv_3 )
\Big\}\Bigg ]CC\go k\bar{k} ,\,
\eee
where\bee\label{CCwgobu}  &&
 \wmv_1{}^\ga:= \gt  z^\ga- (1-\gt )[ (\gn_2+\gn_1\gs_2)p_0   -\gs_1  p_1{} +    \gs_2    p_2{}
]^\ga  \q
\\ \nn
&&\bar \wmv_1{}^\pa:= \bgt  \bz^\pa -(1-\bgt )[
 (\gn_2+\gn_1\gs_1)\bp_0 - \gs_2 \bp_1{} + \gs_1 \bp_2{}  ]^\pa\,,
\\
    \label{W2CCgogt0=}
&& {\wmv_2 }{}^\ga:= -[ ( 1-\gn_1\gs_1)p_{0}  -\gs_1  p _1{}+    \gs_2    p_2{}
 ]^\ga  \q
\\ \nn
&&\bwmv_{2}{}^\pa:= \bgt  \bz^\pa -(1-\bgt )\ga_2[ ( 1-\gn_1\gs_2)\bp_{0}- \gs_2 \bp_1{} + \gs_1 \bp_2{}
 ]^\pa\,\q
\\
\label{w3CCgobgt0=}&&\wmv_3{}^\ga:= \gt  z^\ga- (1-\gt )\ga_2[
 (1-\gn_1\gs_1)p_{0}   - \gs_1  p_1{}+     \gs_2    p_2{}
 ] ^\ga \q
\\ \nn
&&\bwmv_3{}^\pa:= -[  ( 1-\gn_1\gs_2)\bp_{0} - \gs_2 \bp_1{}+ \gs_1 \bp_2{}
 ]^\pa\,
\eee
    with
$\EEE$ \eq{EEE}\,.

 \addtocounter{appendix}{1}
\renewcommand{\theequation}{\Alph{appendix}.\arabic{equation}}
\addtocounter{section}{1} \setcounter{equation}{0}
 \addcontentsline{toc}{section}{\,\,\,\,\,\,\,Appendix C: $\Upsilon^{ \eta}(\go,C,C) $ }

\section*{Appendix C: $\Upsilon^{ \eta}(\go,C,C) $ }
Plugging   $B^\eta_2$  \eqref{B2f} and $W_1$ from \eqref{W1goC} and \eqref{W1Cgo} into the
equation
\be
\dr_x C+[\go,C]_*=-\dr_x B^{\eta }_2-[\go, B^{\eta }_2]_*-[W^{\eta }_1, C]_*+h.c.\,+\ldots
\ee
 after some simple algebra  one finds
using IH  and definitions \eq{ExpO}, \eq{bExp}
\be\label{C2gen}
\dr_x C+[\go,C]_*=
 \Upsilon^\eta_{\go CC}+\Upsilon^\eta_{CC\go}+\Upsilon^\eta_{C\go C}+h.c.\,+\ldots\q
\ee
where
  \bee\label{UpsgoCC}&&
\Upsilon^{\eta}_{\go CC} =   \ff{i\eta}{4}
\int\limits_{\gt\,, \gs(2)  } \ls \tr (\gt)\nabla(  \gs(2) )\nabla(  \gr(2) )    \EE ( \wmv_1  )
   \go C C k   \,,
\\\label{UpsCgoC}&&
\Upsilon^{ \eta}_{C\go  C} =   \ff{i \eta}{4}
\int\limits_{\gt\,, \gs(2)  } \ls \tr (\gt)\nabla(  \gs(2) )\nabla(  \gr(2) ) \EE ( \wmv_2  )
   \go C C k   \,,
\\\label{UpsCCgo}&&
\Upsilon^{ \eta}_{CC\go } =   \ff{i \eta}{4}
\int\limits_{\gt\,, \gs(2)  } \ls \tr (\gt)\nabla(  \gs(2) )\nabla(  \gr(2) )  \EE ( \wmv_3  )
  C   C  \go k   \,
 \eee
with
\bee\label{muCC} \label{kubikUpcgoCC} &&\wmv_1= - (\gr_1+\gs_1\gr_2 ) p_0-\gs_1  p _1+\gs_2 p_2  \q \\
&&\wmv_2=  (-\gs_1\gr_1  +\gs_2\gr_2 ) p_0-\gs_1  p _1+\gs_2 p_2  \q
  \label{kubikUpcCgoC}\\
&&\wmv_3= \, (\gr_2+\gr_1\gs_2 ) p_0-\gs_1  p _1+\gs_2 p_2
  \label{kubikUpcCCgo}\,.\eee

Complex conjugated vertices $\Upsilon^{\bar \eta}$ are analogous.

\addtocounter{appendix}{1}
\renewcommand{\theequation}{\Alph{appendix}.\arabic{equation}}
\addtocounter{section}{1} \setcounter{equation}{0}
 \renewcommand{\thesection}{\Alph{appendix}.}

\section*{Appendix D: Solving for    $\Upsilon^{\eta\bar\eta}_{\go CCC} $ in detail}
\label{UpswCCCDET}
 \addcontentsline{toc}{section}{\,\,\,\,\,\,\,Appendix D: Solving for moderately non-local   $\Upsilon^{\eta\bar\eta}_{\go CCC} $ in detail}
Details of extraction  of  $\Upsilon_{C\go CC}$ \eq{Upsilon reswCCC} from  Eq.~\eq{UwCCC}
are presented in Sections {\bf D.1}-{\bf D.6}.

\subsection*{ D.1\quad $W^{\eta\bar\eta}_2|_{\go CC}*C$}\label{W2C}
\addcontentsline{toc}{subsection}{D.1\quad $W^{\eta\bar\eta}_2|_{\go CC}*C$}

Taking into account   $W_2 |_{\go CC}$
\eq{W2goCCbound}
  along with \eq{w1goCCbu}-\eq{w3goCCbgt0=} one obtains
    \bee\nn&&-
 W^{\eta\bar{\eta}}_2 {}_{\go C{C} }{}*C  =   \ff{\eta\bar{\eta}}{16}
\int\limits_{\gt\, \bgt
 \gs(2)   \gn(2)  }
\ls\nabla(\gs(2))\nabla(\gn(2))
  \times \label{W2goCC*Cbulk}
\Bigg[ -\Box (\gt,\bgt)  \EEE (\wmv_1  | \bar  \wmv_1 )
 \\ \label{W2goCC*Cbound}&&\ls\!+\!  \int\limits_{  \ga(2)}\!\!
 \nabla(\ga(2))  \Big\{
\tr (\gt)l(\bgt) \EEE (\wmv_2  | \bar  \wmv_2 )
\!+\!\tr (\bgt)l(\gt)  \EEE (\wmv_3  | \bar  \wmv_3 )
\Big\}\Bigg ]\go CCC k\bar{k} \,,
\eee
 \bee\label{W2new=*C}    &&
 \wmv_1{} ^\ga  = \gt  z^\ga- (1-\gt )[-(\gn_1+\gn_2\gs_1)p_0  -\gs_1  p _1{} +    \gs_2    p_2{}
+p_3 ]^\ga  \q
\\ \nn
&&\bwmv_1{} ^\pa = \bgt  \bz^\pa -(1-\bgt )[-(\gn_1+\gn_2\gs_2)\bp_0- \gs_2 \bp_1 + \gs_1 \bp_2
+\bp_3 ]^\pa\,,
\\
\label{bwS2polu}
&&\wmv_2{}^\ga:= \gt  z^\ga- (1-\gt )(
\ga_1[ -(\gn_1+\gn_2\gs_1)p_0   -\gs_1  p _1{} +    \gs_2    p_2{}
 ] +p_3{} )^\ga \q
\\ \nn
&&\bwmv_2{}^\pa:= -[- (\gn_1+\gn_2\gs_2)\bp_0 - \gs_2 \bp_1{}+ \gs_1 \bp_2{}
+\bp_3{} ]^\pa\,\q
\\
\label{bwS2polub}
&&\wmv_3{}^\ga:= -[-(\gn_1+\gn_2\gs_1)p_0   -\gs_1  p _1{} +    \gs_2    p_2{} +p_3  ]^\ga  \q
\\ \nn
&&\bwmv_3{}^\pa:= \bgt  \bz^\pa- (1-\bgt )\big\{\ga_1[- (\gn_1+\gn_2\gs_2)\bp_0- \gs_2 \bp_1{} + \gs_1 \bp_2{}
]+\bp_3\big\}^\pa \,.
\eee

The 'bulk' term  of  $- W_2 |_{\go CC }*C$ \eq{W2goCC*Cbulk}, that depends  on $\wmv_1, \bwmv_1$ \eq{W2new=*C}
is canceled by the term of  $(\dr_x+\go*)B^{ \eta\bar{\eta}}_3{}|_{blk}$
\eq{DB3bulkOBRATNO} with $\wmv, \bwmv$  \eq{Ugo21w2B3} generated by $\dr(\theta(\gs_2) )$\,.

The 'boundary'   terms of  $ W_2 |_{\go  CC}*C$  \eq{W2goCC*Cbound}
 are considered in Section {\bf D.6}.

One can easily make sure that all terms in
\eq{W2goCC*Cbound}  do  satisfy \eq{nerav} thus being MNL.

\subsection*{ D.2\quad $( {W}^{\bar \eta}_1* B^{  \eta}_2$+${W}^{  \eta}_1*   B^{\bar \eta}_2)|_{\go CCC}$}
\addcontentsline{toc}{subsection}{D.2\quad $( {W}^{\bar \eta}_1* B^{  \eta}_2$+${W}^{  \eta}_1*   B^{\bar \eta}_2)|_{\go CCC}$}
\label{W1B2}
According to
Eqs.~\eq{S1withw}-\eq{wrtB2b},
taking into account
Eqs.~\eq{ExpO},\,\eq{bExp},\,\eq{EEE},\,\eq{nabla} one gets
\bee
\label{W1B2goCCC}
&&-( {W}^{\bar \eta}_1* B^{  \eta}_2+{W}^{  \eta}_1*   B^{\bar \eta}_2)|_{\go CCC}=
-\ff{\eta\bar{\eta}}{16}\!\ls \int\limits_{\gt \bgt\gn(2) \gs(2) }
  \!\!   \!\!  \nabla(\gs(2)) \nabla(\gn(2))
 \Box( \gt,\bgt)
\quad\\ \nn  && \times
  \Big[\EEE(\wmv_1\, |\bwmv_1)-\EEE(\wmv_2\, |\bwmv_2)\Big] \go CCC k\bar{k},
\eee
 \bee\label{wbarW1B2=}
&&\wmv_1{}^\ga = \gt  z- (1-\gt  )[-p_0  -   p _1-\gs_1 p_2  +     \gs_2   p _3  ]^\ga  \q
\\ &&\nn
 \bwmv_1{}^\pa  = \bgt \bz^\pa -(1-\bgt)(-\gn_1 \bp_0+ \bp_2+\bp_3)^\pa\q
\\\label{wW1barB2=}
  &&\wmv_2{}^\ga   = \tau z^\ga - (1-\tau)( -\gn_1p_0+p_2+p_3  )^\ga\q
 \\ \nn
 &&\bwmv_2{}_{  \pa} = \bgt  \bz^\pa -(1-\bgt )
 [-\bp_0 -    \bp_1  - \gs_2 \bp_2{}+\gs_1 \bp_3   ]^\pa\,.
\eee
Note, that the    terms on the \rhs of \eq{W1B2goCCC} will be canceled below  by
 terms of \eq{DB3bulkOBRATNO} with $\wmv, \bwmv$  \eq{Ugo21w1B3}
  generated by $\dr(\theta(\ga_1)\theta(\ga_2))$.

 \subsection*{ D.3\quad  $(\dr_x+\go*)B^{ \eta\bar{\eta}}_3{}|_{blk}|_{\go CCC} $ }
\addcontentsline{toc}{subsection}{D.3\quad  $(\go*+d)B^{ \eta\bar{\eta}}_3{}|_{blk}|_{\go CCC} $}
\label{SDB3bulk}
Using that $\dr\EEE=0$ one can see that  $ B_3^{\eta\bar \eta}$  \eq{B3bulk}  yields
  \bee \ls&& -   (\dr_x+\go*)B_3^{\eta\bar \eta}|_{blk}|_{\go CCC}=
  \label{DB3bulk}
-   \ff{ \eta\bar{\eta}}{16} \! \! \int\limits_{\gt \bgt  \ga(2)  \gs(2) \gn(2)}\ls\ls\Big( \dr\Big\{
  \Box(\gt,\bgt)
 \nabla(\ga(2))  \nabla(\gs(2))
  \nabla(\gn(2))
 \Big\}
   \rule{0pt}{20pt}
  \\ \label{DB3bulkOBRATNO}
  &&-
   \dr\Big\{ \theta(\ga_1)\theta(\ga_2)\theta(\gs_1)\theta(\gs_2)\Big\}\tr(1-\ga_1-\ga_2)
\tr(1-\gs_1-\gs_2)
  \Box(\gt,\bgt) \nabla(\gn(2))
    \\ \label{DB3bulkOBRATNOCO}
  &&-
 \dr\Big\{\Box(\gt,\bgt)\Big\}
 \nabla(\ga(2))  \nabla(\gs(2))\nabla(\gn(2))
 \Big)   \Big[ \EEE(\wmv_1\, |\bwmv_1)
-\EEE(\wmv_2\, |\bwmv_2)\Big]\go CCC k\bar{k}   \rule{0pt}{20pt}
\,,  \eee where
  \bee\label{Ugo21w1B3}
  &&\wmv_1^{ \ga}:= \gt   z{}^\ga- (1-\gt  )[-(\gn_2+\gn_1\ga_2) p_0
  - \ga_2  (p _1{}+ p_2) + (1-\ga_2\gs_1)(p_3{}+ p_2) ]^\ga  \q
 \\ \nn
&&\bwmv_1^{ \pa}:= \bgt   \bz{}^\pa -(1-\bgt  )[-(\gn_2+\gn_1\ga_1)\bp_0-  \ga_1 (\bp_1{}+\bp_2)
+(1-\ga_1\gs_1)(\bp_3{}+\bp_2) ]^\pa\,,
\\ \label{Ugo21w2B3}
  &&\wmv_2^{ \ga}:= \gt   z{}^\ga- (1-\gt  )[-( 1+\gn_1(  -\ga_2\gs_2))p_0
  -( 1-\ga_2\gs_2) (p _1{}+ p_2) +  \ga_2 (p_3{}+ p_2) ]^\ga  \q
 \\ \nn
&&\bwmv_2^{ \pa}:= \bgt   \bz{}^\pa -(1-\bgt  )[-( 1+\gn_1(  -\ga_1\gs_2))\bp_0
 -(1-\ga_1\gs_2) (\bp_1{}+\bp_2)
  + \ga_1(\bp_3{}+\bp_2) ]^\pa\,.
 \eee
One can see that nontrivial
'bulk' terms of  \eq{DB3bulkOBRATNO} either cancel
  each other or cancel $- {W}^{  \bar{\eta}}_1* B^{ \eta }_2  |_{\go CCC}$ \eq{wbarW1B2=},
 $-W^{ \eta }_1* {B}^{  \bar{\eta}}_2 $
  \eq{wW1barB2=} and the 'bulk'  term of $-W^{ \eta\bar{\eta}}_2*C|_{\go CCC}$ \eq{W2new=*C}.
Hence all 'bulk' terms on the \rhs of \eq{dxC=} in the sector under consideration   vanish.
The next step is to consider   'boundary' terms.

Note that since $B^{ \eta\bar{\eta}}_3$
satisfies \eq{nerav},  $\D B^{ \eta\bar{\eta}}_3(\go,C,C,C)$  satisfies it as well.
\subsection*{ D.4\quad $(\dr_x+\go*)B^{ \eta\bar{\eta}}_3{}|_{bnd}{}|_{\go CCC} $   }
\addcontentsline{toc}{subsection}{D.4\quad $(d_x+\go*)B^{ \eta\bar{\eta}}_3{}|_{bnd}{}|_{\go CCC} $}
\label{DB3boundyields1}

From \eq{B3bound} along with    \eq{31w3B3===},   \eq{31w4B3==}\, it follows\bee  \label{B3bound=}
  &&-(\dr_x+\go*)B^{ \eta\bar{\eta}}_3{}|_{bnd}{}|_{\go CCC}=\ff{\eta\bar{\eta}}{16}\!\!
  \int\limits_{\gt \bgt\gn(2) \ga(2) \gx(3)}
  \!\!   \!\!\dr\Big\{ \mu_1\times
\\ \nn  &&\Big[\tr(\bgt)l(\gt)  \EEE(\wmv_3\, |\bwmv_3)
-\tr(\gt)l(\bgt)  \EEE(\wmv_4\, |\bwmv_4)\Big]\Big\}  \go CCC k\bar{k}
 \\ \label{ObratnoB3bound} &&-\ff{\eta\bar{\eta}}{16}\!\! \int\limits_{\gt \bgt\gn(2) \ga(2) \gx(3)}
  \!\!   \!\! \tr(1-\ga_1-\ga_2)  \nabla( \gn(2)) \tr(1-\gx_1-\gx_2-\gx_3)
     \times
\\ \nn  && \dr\Big\{\theta(\gx_1)\theta(\gx_2)\theta(\gx_3)
     \theta(\ga_1)\theta(\ga_2)\Big\}
  \times
\\ \nn  &&\Big[\tr(\bgt) l(\gt)    \EEE(\wmv_3\, |\bwmv_3) \
 -\tr(\gt) l(\bgt)  \EEE(\wmv_4\, |\bwmv_4)\Big]  \go CCC k\bar{k}
 \\ \label{COHOMDB3} &&
-\ff{\eta\bar{\eta}}{16}\!\! \int\limits_{\gt \bgt\gn(2) \ga(2) \gx(3)}
  \!\!   \!\!\tr(\gt)\tr(\bgt)  \mu_1
\Big[\EEE(\wmv_3\, |\bwmv_3)
+\EEE(\wmv_4\, |\bwmv_4)\Big] \go CCC k\bar{k}  \eee
with
 $\mu_1 =  \nabla(\ga(2))\nabla(\gn(2)) \nabla(\gx(3))$ and
 \bee\label{31w3B3r} &&\wmv_3^{ \ga}:= \gt   z{}^\ga- (1-\gt  )
 \left [ -(\gn_2+\gn_1\{\ga_1  \gx_2 ({1-\gx_1})^{-1}+\gx_3\}) p_0\right.
  \\ \nn&&
 \left.  -  \{   \ga_1  \gx_2 ({1-\gx_1})^{-1}+\gx_3
   \} (p _1{}+ p_2) +  \{{\ga_2 \gx_2 }({1-\gx_3})^{-1}
  +\gx_1   \}(p_3{}+ p_2) \right]^\ga  \q
 \\ \nn
&&\bwmv_3^{ \pa}:= -[-(\gn_2+\gn_1(1-\gx_3)) \bp_0  -(1-\gx_3) (\bp_1{}+\bp_2) + (1-\gx_1)(\bp_3{}+\bp_2) ]^\pa\,,
 \\
  \label{31w4B3r}
  &&
 \wmv_4^{ \ga}:=-[ -(\gn_2+\gn_1(1-\gx_3)) p_0  - ( 1-\gx_3)  (p _1{}+ p_2)+ (1-\gx_1)(p_3{}+ p_2) ]^\ga  \q
 \\ \nn
&&\bwmv_4^{ \pa}:= \bgt   \bz{}^\pa -(1-\bgt  )
[ -(\gn_2+\gn_1\{ \ga_1  \gx_2  (1-\gx_1)^{-1}+\gx_3 \} ) \bp_0
\\ &&\nn-
 \{   \ga_1  \gx_2  (1-\gx_1)^{-1} +\gx_3\} (\bp_1{}+\bp_2)
 +\left\{\ga_2 \gx_2(1-\gx_3)^{-1} + \gx_1\right\}(\bp_3{}+\bp_2) ]^\pa\,.
 \eee

  One can see that the terms
  of \eq{ObratnoB3bound}  generated by
$
\dr\big\{
     \theta(\ga_1)\theta(\ga_2)\big\}$
cancel against
the respective 'boundary' terms of \eq{DB3bulkOBRATNOCO}  by   \eq{ratioksibgt}-like changes of variables.
The rest nontrivial terms of \eq{ObratnoB3bound}, namely $\dr \theta(\gx_1)  $-dependent ones,
   will be considered in  Section {\bf D.6}. %\ref{LastgoCCCHOM}.

Note, that cohomology terms \eq{COHOMDB3} are represented in Eq.~\eq{Upsilon reswCCC}
of Section  \ref{wCCC}.

\subsection*{ D.5\quad$\big(\dr_x B^{\bar\eta}_2 +\dr_x B^{ \eta}_2\big)   |_{\go CCC}$}
 \addcontentsline{toc}{subsection}{D.5\quad $\big(d_x B^{\bar\eta}_2 +d_x B^{ \eta}_2\big)   |_{\go CCC}$}

\label{dxbB2wCCC}
 By virtue of \eq{C2gen}, taking into account \eq{UpsgoCC} and its  conjugated, one obtains from \eq{B2f} and
 \eq{bB2f}\bee\label{dxdB2CgoCC1}
 &&-\big(\dr_x B^{\bar\eta}_2 +\dr_x B^{ \eta}_2\big) |_{\go CCC}=
\ff{\eta\bar{\eta}}{16}\!\ls \int\limits_{\gt \bgt\gn(2) \gs(2) }
  \!\!   \!\ls\nabla(\ga(2))\nabla(\gs(2))\nabla(\gn(2))     \\ \nn  &&
      \Big[- \tr(\bgt )l(\gt)\EEE(\wmv_1\, |\bwmv_1)
  +\tr(\gt )l(\bgt)\EEE(\wmv_2\, |\bwmv_2)\Big] \go CCC k\bar{k}\,,
\eee
\bee\label{dxbB2}
 &&\wmv_1{}^\ga =
-[- (1-\gs_2\gn_2  ) p_0 -\gs_2  p _1 + \gs_1 p_2 +p_3 ]{}^\ga   \\ \nn
&& \bwmv_1{}{}^\pa  =  \bgt \bz^\pa -(1- \bgt)[-\ga_1(\bp_0+\bp_1+\bp_2)
+ \ga_2\bp_3   ]^\pa
  \,,
\\
\label{dxB2}
 &&\wmv_2{}^\ga = \tau z^\ga -(1-\gt)
[  -\ga_1 (p_0+ p_1+ p_2)+\ga_2 p_3{}]^\ga \\ \nn
&& \bwmv_2{} {}^\pa  =
[- (1-\gs_2\gn_2) \bp_0  -\gs_2  \bp _1+\gs_1 \bp_2+ \bp_3
     ]^\pa \,.
 \eee
These terms are considered in Section {\bf D.6}.
One  can easily make sure that   $\big(\dr_x B^{\bar\eta}_2 +\dr_x B^{ \eta}_2\big)  |_{\go CCC}$ \eq{dxdB2CgoCC1} satisfies \eq{nerav}, thus being MNL.

\subsection*{D.6 \quad The rest cohomology terms}
 \addcontentsline{toc}{subsection}{D.6 \quad Rest cohomology terms}

 \label{LastgoCCCHOM}
 Here we consider the rest
  'boundary' terms at $\bgt=0$ dependent on $\wmv,\bwmv$ of the form \eq{bwS2polu},
  \eq{31w3B3r} at  $ \gx_1=0$
 and \eq{dxB2},
 as well as the 'boundary' terms at $\gt=0$ dependent on $\wmv,\bwmv$ of the form
  \eq{bwS2polub}, \eq{31w4B3r} at  $ \gx_1=0$
and \eq{dxbB2}.

To obtain rest cohomology terms from those with $\bgt=0$ consider
\bee\label{lastUnifbgt0}\ls  &&  \ff{\eta\bar{\eta}}{16}\!\!
 \int\limits_{\gt \bgt\gn(2) \ga(2)\gb(2) \gs(2) }
\!\!  \!\!   \!\!\dr
\Big\{\tr(\bgt) l( \gt )\nabla(\ga(2))\nabla(\gs(2))\nabla(\gn(2))\nabla(\gb(2))
   \EEE(\wmv |\bwmv)  \go CCC k\bar{k}
\Big\} \eee
with
\bee \label{wbwlastbgt0}
 &&  \wmv^\ga=\tau z^\ga -(1-\gt)\Big[ \ga_2\big\{
\gb_1[ -( 1-\gn_2\gs_2)p_0  -\gs_1  p _1{}+    \gs_2    p_2{}
  ] +p_3{} \big\}
\\ \nn&&   -\gb_1\ga_1 (  p _1+ p_2 + p_0)   +\ga_1\gb_2   p_3{} \Big]^\ga\\ \nn&&
 \bwmv^\pa=-[-(\gn_1+\gn_2 \gs_2) \bp_0  - \gs_2 (\bp_1{}+\bp_2) +  (\bp_3{}+\bp_2) ]^\pa\,.
\eee
  Expression \eq{lastUnifbgt0} is
 in the $d Z$-independent sector and hence   gives zero  as an integral of an exact form.
 (Recall, that in this sector  we discard $d Z$-dependent weak terms.)

Differentiation  in \eq{lastUnifbgt0}  gives  the cohomology term (with a sign "-")
of \eq{Upsilon reswCCC}
that depends on   $\Omega_3\,,\bar \Omega_3$ \eq{31w3B3COH}  along with all
the rest
  'boundary' terms with $\bgt=0$.
Namely, the term
$(\sim \dr(\theta(\gb_2)))$  equals to the term in \eq{ObratnoB3bound} that depends on
$\wmv_3, \bwmv_3$  \eq{31w3B3r}   at $\gx_1=0$,
 the term ($\sim \dr(\theta(\ga_1))$)  equals to a part of $-W^{ \eta\bar \eta}_2*C|_{\go CCC}$
\eq{W2goCC*Cbound} that depends on
$\wmv , \bwmv $ of the form \eq{bwS2polu} while that   ($\sim \dr(\theta(\ga_2))$) equals
 to the part of $-d_x B^{ \eta}_2  |_{\go CCC}$ \eq{dxdB2CgoCC1} that depends on
$\wmv , \bwmv $  \eq{dxB2}.
Note that the term $\sim \dr(\theta(\gb_1))$ is  weak since $\wmv^\ga|_{\gb_1=0}=\tau z^\ga -(1-\gt)p_3^\ga$.

Analogously, differentiation in the following expression
\bee\label{lastUnifgt0} &&- \ff{\eta\bar{\eta}}{16}\ls \int\limits_{\gt \bgt\gn(2) \ga(2)\gb(2) \gs(2) }
\ls  \ls d
\Big\{\tr(\gt) l( \bgt ) \nabla(\ga(2))\nabla(\gs(2))\nabla(\gn(2))\nabla(\gb(2))
   \EEE(\wmv |\bwmv)  \go CCC k\bar{k}
\Big\}\quad
\eee
with
\bee \label{wbwlastgt0} &&
  \wmv^\ga=-[-(\gn_1+\gn_2 \gs_2) p_0  - \gs_2 ( p_1{}+ p_2) +  ( p_3{}+ p_2) ]^\ga
\,,\\ \nn
 &&  \bwmv^\pa=\bgt \bz^\pa -(1-\bgt)\Big[ \ga_2\big\{
\gb_1[ -( 1-\gn_2\gs_2)\bp_0  -\gs_1  \bp _1{}+    \gs_2    \bp_2{}
  ] +\bp_3{}\big\}
  \\ \nn
&&   -\gb_1\ga_1 (  \bp _1+ \bp_2 + \bp_0)   +\ga_1\gb_2   \bp_3{} \Big]^\pa\eee
gives all the rest 'boundary' terms with $\gt=0$ plus a cohomological one.
Namely we obtain
    the  cohomology term (with a sign "-")
of \eq{Upsilon reswCCC},
that depends on $\wmv_4,\bwmv_4$ \eq{31w4B3rCOH},
along with the  term
   of $ (\dr_x+\go*)B^{ \eta\bar{\eta}}_3{}|_{bnd}{}|_{\go CCC}$
   that depends on  $\wmv_4 , \bwmv_4 $
\eq{31w4B3r} at $\gx_1=0$, the  term of  $W_2*C|_{\go CCC}$,
 that depends on
$\wmv_3 , \bwmv_3 $ \eq{bwS2polub}
and   $\dr_x B^{\bar\eta}_2|_{\go CCC}$, that depends on $\wmv_1 , \bwmv_1 $ \eq{dxbB2}.
 Note that the expressions \eq{wbwlastgt0} result from the application of MNL preserving IH
 to $-W_2*C|_{\go CCC}$ and  $-\dr_x B^{\bar\eta}_2|_{\go CCC}$, which are MNL  (see \eq{bwS2polub}, \eq{dxbB2}).

 Thus  all cohomological terms in the sector $\go CCC$ are extracted    from Eqs.~\eq{COHOMDB3},
\eq{lastUnifbgt0} and \eq{lastUnifgt0} yielding
$\Upsilon^{\eta\bar\eta}|_{\go CCC}$ \eq{Upsilon reswCCC}.

\addtocounter{appendix}{1}
\renewcommand{\theequation}{\Alph{appendix}.\arabic{equation}}
\addtocounter{section}{1} \setcounter{equation}{0}
 \renewcommand{\thesection}{\Alph{appendix}.}
  \renewcommand{\thesubsection}{\Alph{appendix}.\arabic{subsection}}

 \addcontentsline{toc}{section}{\,\,\,\,\,\,\,Appendix E: Solving for moderately non-local
 $\Upsilon^{\eta\bar\eta}_{C\go CC} $ in detail}

 \section*{Appendix E: Solving for moderately non-local   $\Upsilon^{\eta\bar\eta}_{C\go CC} $}
\label{UpsCwCCDET}

Details of derivation of   $\Upsilon_{C\go CC}$ \eq{ResultCgoCCb1} from  Eq.~\eq{UCwCC} are presented in   Sections
{\bf E.1}-{\bf E.8}.

\setcounter{section}{0}
\addcontentsline{toc}{subsection}{E.1 \quad $C*(W_2|_{ \go CC})$ }
 \subsection*{E.1 \quad $  C*(W_2|_{ \go CC})$}

 \label{CW2wCC}

Using    $W_2 |_{\go CC}$
\eq{W2goCCbound} along with \eq{w1goCCbu}-\eq{w3goCCbgt0=} one obtains
 \bee\label{CW2CgoCC}&&
 C*( W_2 {}|_{\go C{C} })  =    \ff{\eta\bar{\eta}}{16}
\int\limits_{\gt\,  \bgt  \gs(2)   \gn(2)  } \ls\mu_1
 \Box (\gt,\bgt)   \EEE (\wmv_1|\bwmv_1)
  C\go CCk\bar{k}
\\ \label{CW2CgoCCbound}&&  -\ff{\eta\bar{\eta}}{16}
\int\limits_{\gt\,  \bgt  \gs(2)   \gn(2) \ga(2)} \ls
\mu_1\mu_2   \Big\{
\tr (\gt)l(\bgt) \EEE (\wmv_2|\bwmv_2)
+\tr (\bgt)l(\gt)  \EEE (\wmv_3|\bwmv_3)
\Big\}C\go CCk\bar{k} \,
\eee
with $\mu_1= \nabla(\gs(2))\nabla(\gn(2)) \q
 \mu_2=\nabla(\ga(2))
 \q$
  \bee\label{C*W2new=}   &&
 \wmv_1{}^\ga  = \gt  z^\ga- (1-\gt )[  -(\gn_1+\gn_2\gs_1)p_0 -p_1 -\gs_1  p _2{} +    \gs_2    p_3{}
 ]^\ga  \q
\\ \nn &&\bar{\wmv} _1{}^\pa = \bgt  \bz^\pa -(1-\bgt )[-(\gn_1+\gn_2\gs_2)\bp_0-\bp_1{}- \gs_2 \bp_2{} + \gs_1 \bp_3{}
   ]^\pa\,,
\\  \label{W2goCCw2}
&&\wmv_2{}^\ga = -[ -(\gn_1+\gn_2\gs_1)p_0-p_1  -\gs_1  p_2{} +    \gs_2    p_3{} ]^\ga  \q
\\ \nn%\label{bwS2}
&&\bwmv_2{ }^\pa= \bgt  \bz^\pa -(1-\bgt ) [-\ga_1(\gn_1+\gn_2\gs_2)\bp_0
-\bp_1- \ga_1\gs_2 \bp_2{}
+ \ga_1\gs_1 \bp_3{}
]^\pa\,,
\\
\label{W2goCCw3}&&\wmv_3{ }^\ga = \gt  z^\ga- (1-\gt )[ -\ga_1(\gn_1+\gn_2\gs_1)p_0-p_1   -\ga_1\gs_1  p_2{}+    \ga_1\gs_2    p_3{}
]^\ga  \q
\\ \nn%\label{bwS2}
&&\bwmv_3{ }^\pa = -[
- (\gn_1+\gn_2\gs_2)\bp_0-\bp_1- \gs_2 \bp_2{} + \gs_1 \bp_3  ]^\pa\,.
\eee
Note that the term in $ C*(W_2|_{ \go CC})$, that depends on $\wmv_1,\,\bwmv_1$ \eq{C*W2new=},
is cancelled  in Section  {\bf E.4}
The 'boundary' terms
 dependent on $\wmv_2,\,\bwmv_2$ \eq{W2goCCw2} and  $\wmv_3,\,\bwmv_3$  \eq{W2goCCw3}
 are considered in  Section {\bf E.8}  .

  \subsection*{E.2 \quad $W_2|_{C\go C}*C$ }
\addcontentsline{toc}{subsection}{E.2 \quad $W_2|_{C\go C}*C$ ${}$ }
\label{W2CgoCC}

 Using $W_2 |_{C\go C}$   \eq{W2CgoCbound} along with \eq{OW2CgoC0}-\eq{PW2CgoC131=4} one obtains
\bee\label{sumW2C}  &&
- W_2 {}_{C\go C}*C   = -   \ff{\eta\bar{\eta}}{16}
\ls\ls\int\limits_{\gt\,, \bgt\,, \gr(2)  \,,\gs(2)  }
\ls\ls \mu_1 \Box (\gt,\bgt)  \Big\{ \EEE (\wmv_0|\bwmv_0)+\EEE (\wmv_1|\bwmv_1)
+\EEE (\wmv_2|\bwmv_2)\Big\}C\go CC k\bar{k}  \,,
\eee
     \bee\label{W2goCCboundgt}&&  - \ff{\eta\bar{\eta}}{16}
\int\limits_{\gt\,, \bgt\,, \gn(2)  \,,\gs(2) \,,  \gx(2)}\ls\ls\ls
\mu_1\mu_2 \Big\{
\tr (\gt)l(\bgt)  \EEE (\wmv_3|\bwmv_3)
-\tr (\bgt)l(\gt)  \EEE (\wmv_4|\bwmv_4)
\Big\}C\go CC k\bar{k} \,
\eee
 with
 $\mu_1= \nabla(\gs(2))\nabla(\gn(2)) \q
 \mu_2=\nabla(\ga(2))$,
\bee \label{COW2CgoC0}  &&\wmv_0{}^\ga = \gt  z^\ga- (1-\gt )[ (-\gs_1 \gn_1+\gs_2 \gn_2)p_0 -\gs_1  p _1{} +    \gs_2    p_2{}  +p_3]^\ga   \q
\\ \nn
&&\bwmv_0{}^\pa = \bgt  \bz^\pa -(1-\bgt )[(-\gs_2 \gn_1+\gs_1 \gn_2)  \bp_{ 0}- \gs_2 \bp_1{}
+ \gs_1 \bp_2{}   +\bp_3]^\pa
\q
\\ \label{COW2CgoC1} &&\wmv_1{}^\ga = \gt  z^\ga- (1-\gt )[ ( -\gs_1 \gn_2-  \gn_1) p_0 -   p _1{}   +p_3]^\ga   \q
\\ \nn
&&\bwmv_1{}^\pa= \bgt  \bz_1{}^\pa -(1-\bgt )[( \gs_1 \gn_1+  \gn_2) \bp_{ 0}+\bp_2{}   +\bp_3]^\pa \q
\\
\label{COW2CgoC2}
 &&\wmv_2{}^\ga= \gt  z^\ga- (1-\gt )[ (\gs_2 \gn_1+ \gn_2) p_0 +   p _2{}   +p_3]^\ga   \q
\\ \nn
&&\bwmv_2{}^\pa= \bgt  \bz^\pa -(1-\bgt )[-( \gs_2 \gn_2+  \gn_1) \bp_{ 0}-\bp_1{}   +\bp_3]^\pa
 \q
\\
\label{COW2CgoC3}  && \wmv_{3}{}^\ga= -
( [\gn_2\gs_2 -\gn_1 \gs_1 ] p_0-\gs_1 p_1+\gs_2 p_2 +p_3)^\ga
\q\\ \nn
&&\bwmv_{3}{} ^\pa= \bgt  \bz^\pa- (1-\bgt )
\{\gx_1   ([\gn_2\gs_1    -\gn_1\gs_2  ] \bp_0-\gs_2\bp_1+\gs_1\bp_2 )+\bp_3\}^\pa \q
\\
\label{COW2CgoC4}
&& \wmv_{4}{} ^\ga=  \gt  z^\ga- (1-\gt )
\{\gx_1   ([\gn_2\gs_1    -\gn_1\gs_2  ]  p_0-\gs_2 p_1+\gs_1 p_2
 )+p_3\}^\ga \q\\ &&\nn
  \bwmv_{4}{}^\pa= -
( [\gn_2\gs_2 -\gn_1 \gs_1 ] \bp_0-\gs_1 \bp_1+\gs_2 \bp_2 +\bp_3)^\pa
  \,.
 \eee

Note that the   $\wmv_0,\bwmv_0$ dependent term \eq{COW2CgoC0}
 cancels against that proportional to $\dr\big[\theta(\gs_1)\big]$
 of $\dr B_3^{\eta\bar \eta}|_{blk}{}|_{C\go CC}$ \eq{DB3bulkOBRATNOOCwCC}
that   depends on $\wmv , \bwmv $  \eq{2Ugo21w1B3s1}, while the
   terms   dependent on $\wmv, \bwmv$\eq{COW2CgoC1} and \eq{COW2CgoC2} are considered in Section  {\bf E.7}. The terms
dependent on $\wmv, \bwmv$ \eq{COW2CgoC3} and \eq{COW2CgoC4} are considered in Section  {\bf E.8}.

    \subsection*{E.3 \quad $\left( {W}^{\bar\eta}_1* B^{ \eta}_2 +W^{ \eta}_1*  {B}^{\bar\eta}_2\right)|_{C\go CC}$}
\addcontentsline{toc}{subsection}{E.3 \quad $\left( {W}^{\bar\eta}_1* B^{ \eta}_2 +W^{ \eta}_1*  {B}^{\bar\eta}_2\right)|_{C\go CC}$}
 \label{W1B2+}
According to \eq{W1Cgo}, \eq{B2f} and \eq{bB2f}\be
\label{W1B2gCgoCC}
-\big( {W}^{\bar\eta}_1* B^{ \eta}_2 +W^{ \eta}_1*  {B}^{\bar\eta}_2\big) |_{C\go CC}=
-\ff{\eta\bar{\eta}}{16}\!\ls \int\limits_{\gt \bgt\gr(2) \gs(2) }
  \!\!   \! \mu \,\,
 \Box( \gt,\bgt)
   \Big[\EEE(\wmv_1|\bwmv_1)- \EEE(\wmv_2|\bwmv_2)\Big] C\go CC k\bar{k},
\ee
with
 $\mu= \nabla(\gs(2))\nabla(\gr(2)) \q
 $
 \bee\label{wbarW1B2CgoCC}
  &&\wmv_1{}^\ga  = \gt  z^\ga- (1-\gt  )[  -  (p_0+p _1 ) - \gr_1 p_2   +\gr_2 p _3 ]{}^\ga  \q
\\ && \nn \bwmv_1{}^\ga = \bgt \bz^\pa -(1-\bgt)( \gs_1 \bp_0+\bp_2+\bp_3)^\pa\q
\\
\label{wW1barB2=CgoCC}
  &&\wmv_2{}^\ga := \tau z^\ga - (1-\tau)(\gs_1  p_0 +p_2+p_3 )^\ga\q
 \\\nn &&\bwmv_2{}_{  \,\pa}  = \bgt  \bz -(1-\bgt)[-   (\bp_0+\bp_2+\bp_1){}
 + \gr_1 (\bp_2{}+\bp_3)   ]^\pa\,.
\eee

The term  \eq{W1B2gCgoCC} is considered in   Section  {\bf E.7}.

\subsection*{E.4 \quad $\dr_x B_3{}|_{blk} |_{C\go CC}$  }  %OKT22}
\addcontentsline{toc}{subsection}{E.4 \quad $\dr_x B_3{}|_{blk} |_{C\go CC}$ }
\label{DB3bulkCwCC}
Using that $\dr\EEE=0$ one can see that Eq.~\eq{B3bulk}
  yields
\bee &&    - \dr B_3^{\eta\bar \eta}|_{blk}{}|_{C\go CC}=
  \label{DB3bulkOCwCC}
-    \ff{\eta\bar{\eta}}{16}\!\! \int\limits_{\gt \bgt\, \gn(2)  \gs(2)  \gx(2)}\!\ls \Big( \dr\Big\{
  \Box(\gt,\bgt) \mu_1 \mu_2\Big\}
   \rule{0pt}{20pt}
  \\ \label{DB3bulkOBRATNOOCwCC}
  &&-
   \dr\Big\{\mu_1\Big\}
 \Box(\gt,\bgt) \mu_2
    \\ \label{DB3bulkOBRATNOCOCwCCpolu}
  &&-
 \dr\Big\{\Box(\gt,\bgt)\Big\}
 \mu_1 \mu_2
\Big)\Big[ \EEE(\wmv_1|\bwmv_1)
-\EEE(\wmv_2|\bwmv_2)\Big]\go CCC k\bar{k}   \rule{0pt}{20pt}
\,,  \eee
with $\mu_1= \nabla(\gs(2))\nabla(\gn(2))\,,
 \mu_2=\nabla(\gx(2))$\q
  \bee\label{2Ugo21w1B30}
 &&\wmv_1^{ \ga}:= \gt   z{}^\ga- (1-\gt  )[ (\gx_1(1-\gn_2\gs_1)- \gn_2) p_0 - \gn_2  (p _1{}+ p_2)
 + (1-\gn_2\gs_1)(p_3{}+ p_2) ]^\ga  \q
 \\ \nn% \label{221bw1B3}
&&\bwmv_1^{ \pa}:= \bgt   \bz{}^\pa -(1-\bgt  )[ (\gx_1(1-\gn_1\gs_1)- \gn_1)\bp_0
-  \gn_1 (\bp_1{}+\bp_2)    +(1-\gn_1\gs_1)(\bp_3{}+\bp_2)]^\pa\q
\\ \label{2Ugo21w2B3}
  &&\wmv_2^{ \ga}:= \gt   z{}^\ga- (1-\gt  )[ ( -( 1-\gn_2\gs_2)+\gx_1 \gn_2)p_0
-( 1-\gn_2\gs_2) (p _1{}+ p_2) +  \gn_2 (p_3{}+ p_2) ]^\ga  \q\quad
\\&&\nn% \label{221bw2B3}
\bwmv_2^{ \pa}:= \bgt   \bz{}^\pa -(1-\bgt  )[ ( -(1-\gn_1\gs_2) +\gx_1 \gn_1 )\bp_0
-(1-\gn_1\gs_2) (\bp_1{}+\bp_2)
  + \gn_1(\bp_3{}+\bp_2) ]^\pa\,.
 \eee
Non-zero terms of \eq{DB3bulkOBRATNOOCwCC} are those that depend on
   \bee\label{12Ugo21w1B3}
  &&\wmv_1^{ \ga}|_{\gn_1=0}= \gt   z{}^\ga- (1-\gt  )[ (\gx_1 \gs_2- 1) p_0 -
 (p _1{}+ p_2) +  \gs_2(p_3{}+ p_2) ]^\ga  \q
 \\ \nn% \label{221bw1B3}
&&\bwmv_1^{ \pa}|_{\gn_1=0}= \bgt   \bz{}^\pa -(1-\bgt  )[ (\gx_1 )\bp_0    +
(\bp_3{}+\bp_2) ]^\pa\,\,,
\\
\label{2Ugo21w1B3}
  &&\wmv_1^{ \ga}|_{\gn_2=0}= \gt   z{}^\ga- (1-\gt  )[ (\gx_1 ) p_0  + (1 )(p_3{}+ p_2) ]^\ga  \q
 \\ \nn% \label{221bw1B3}
&&\bwmv_1^{ \pa}|_{\gn_2=0}= \bgt   \bz{}^\pa -(1-\bgt  )[ (\gx_1   \gs_2-  1)\bp_0-    (\bp_1{}+\bp_2)
 +\gs_2 (\bp_3{}+\bp_2) ]^\pa\,,
\\
\label{2Ugo21w1B3s1}
  &&\wmv_1^{ \ga}|_{\gs_1=0}= \gt   z{}^\ga- (1-\gt  )[ (\gx_1 - \gn_2) p_0 - \gn_2  (p _1{}+ p_2)
  +  (p_3{}+ p_2) ]^\ga  \q
 \\ \nn &&\bwmv_1^{ \pa}|_{\gs_1=0}= \bgt   \bz{}^\pa -(1-\bgt  )[ (\gx_1 - \gn_1)\bp_0-  \gn_1 (\bp_1{}+\bp_2)
+ (\bp_3{}+\bp_2) ]^\pa\,,
\\
 \label{2Ugo21w2B3s2}
  &&\wmv_2^{ \ga}|_{\gs_2=0}= \gt   z{}^\ga- (1-\gt  )[ ( - 1+\gx_1 \gn_2)p_0
-  (p _1{}+ p_2) +  \gn_2 (p_3{}+ p_2)]^\ga  \q
 \\ \nn &&\bwmv_2^{ \pa}|_{\gs_2=0}= \bgt   \bz{}^\pa -(1-\bgt  )[ ( - 1  +\gx_1\gn_1)\bp_0
-  (\bp_1{}+\bp_2)
  + \gn_1(\bp_3{}+\bp_2) ]^\pa\,.
 \eee
Note, that the sum of the terms on the \rhs of \eq{DB3bulkOBRATNOOCwCC} dependent on $ (\wmv_1\,, \bwmv_1 )|_{\gs_2=0}$ and
 $ (\wmv_2\,, \bwmv_2 )|_{\gs_2=0}$ gives zero.
The   term  of   \eq{DB3bulkOBRATNOOCwCC}, that depends  on
  $\wmv_2 |_{\gs_2=0}\,, \,\bwmv_2 |_{\gs_2=0}$
 \eq{2Ugo21w2B3s2} cancels the term of
 $C*(W_2|_{ \go CC})$ \eq{C*W2new=}, while the term
 that depends  on
  $\wmv_1 |_{\gs_1=0}\,, \,\bwmv_1 |_{\gs_1=0}$
   \eq{2Ugo21w1B3s1} cancels
 the   term  of  $-W_2 {}_{C\go C}*C$ \eq{sumW2C}, that depends  on
  $\wmv , \bwmv$ \eq{COW2CgoC0}.

 The  non-zero 'boundary'  terms  of   \eq{DB3bulkOBRATNOCOCwCCpolu} are cancelled by the respective
 terms of \eq{2ObratnoB3bound}
 from
$\dr\big\{
     \theta(\ga_1)\theta(\ga_2)\big\}$ as can be seen with the help
      of the \eq{ratioksibgt}-like changes of variables.

   The rest non-zero $ (\wmv_1 \,, \,\bwmv_1)|_{\gn_{1,2}=0}$-dependent terms of \eq{DB3bulkOBRATNOOCwCC}
      associated with \eq{12Ugo21w1B3}, \eq{2Ugo21w1B3} are considered
   in Section  {\bf E.7}.

\subsection*{E.5 \quad    $ \dr_x B_3{}|_{bnd}|_{C\go CC} $  }
\addcontentsline{toc}{subsection}{E.5 \quad    $ \dr_x B_3{}|_{bnd}|_{C\go CC} $   }\label{DB3boundCwCC}

From \eq{B3bound} along with    \eq{31w3B3===},   \eq{31w4B3==}\, it follows
 \bee  \label{2DB3bound=}
  &&\dr_x B_3{}|_{bnd}{}|_{C\go CC}=-\ff{\eta\bar{\eta}}{16}\!\!
  \int\limits_{\gt \bgt\gs(2) \gr(2) \gx(3)}
  \!\!   \!\!\dr\Big\{ \nabla(\gr(2))\nabla(\gs(2)) \nabla(\gx(3))
   \times
\\ \nn  &&\Big[\tr(\bgt) l(\gt)  \EEE(\wmv_3|\bwmv_3)
+\tr(\gt) l(\bgt)  \EEE(\wmv_4|\bwmv_4)\Big]\Big\} C \go CC k\bar{k}
 \\ \label{2ObratnoB3bound} &&-\ff{\eta\bar{\eta}}{16}\!\!
\int\limits_{\gt \bgt\gs(2) \gr(2) \gx(3)}
  \!\!   \!\! \tr(1-\gr_1-\gr_2)    \nabla( \gs(2)) \tr(1-\gx_1-\gx_2-\gx_3)
     \times
\\  \nn  && \dr\Big\{\theta(\gx_1)\theta(\gx_2)\theta(\gx_3)
     \theta(\gr_1)\theta(\gr_2)\Big\}
 \\ \nn  &&\Big[\tr(\bgt) l(\gt)    \EEE(\wmv_3|\bwmv_3) \
 -\tr(\gt) l(\bgt)  \EEE(\wmv_4|\bwmv_4)\Big] C \go CC k\bar{k}
 \\ \label{2COHOMDB3} &&
-\ff{\eta\bar{\eta}}{16}\!\!\!\ls \int\limits_{\gt \bgt\gs(2) \gr(2) \gx(3)}
  \!   \ls\ls\tr(\gt)\tr(\bgt) \nabla(\gr(2))\nabla(\gs(2)) \nabla(\gx(3))
 \Big[\EEE(\wmv_3|\bwmv_3)
+\EEE(\wmv_4|\bwmv_4)\Big] C\go CC k\bar{k}\,,  \eee
 \bee\label{CgoCCw3B3r}
 &&\wmv_3{}^{ \ga} = \gt   z{}^\ga- (1-\gt  )
 \left [  (\gn_1\{{\gr_2 \gx_2 }({1-\gx_3})^{-1}
  +\gx_1   \}- \{\gr_1  \gx_2 ({1-\gx_1})^{-1}+\gx_3\}) p_0
 \right. \\ \nn&&
 \left.-  \{   \gr_1  \gx_2 ({1-\gx_1})^{-1}+\gx_3
   \} (p _1{}+ p_2)
 +  \{{\gr_2 \gx_2 }({1-\gx_3})^{-1}
  +\gx_1   \}(p_3{}+ p_2) \right]^\ga  \q
 \\ \nn
&&\bwmv_3{}^{ \pa}:= -[ (\gn_1(1-\gx_1)- (1-\gx_3)) \bp_0  -(1-\gx_3) (\bp_1{}+\bp_2)
+ (1-\gx_1)(\bp_3{}+\bp_2) ]^\pa\,,
 \\
  \label{31w4B3rCwCC}
 &&
 \wmv_4{}^{ \ga}:=-[  (\gn_1(1-\gx_1)-(1-\gx_3)) p_0  - ( 1-\gx_3)  (p _1{}+ p_2)
 + (1-\gx_1)(p_3{}+ p_2) ]^\ga  \q
 \\ \nn
&&\bwmv_4{}^{ \pa} = \bgt   \bz{}^\pa -(1-\bgt  )[  (\gn_1\{\gr_1 \gx_2(1-\gx_3)^{-1} + \gx_1 \}
-\{  \gx_3+\gr_2  \gx_2  (1-\gx_1)^{-1} \} ) \bp_0
 \\ &&\nn- \{  \gx_3+\gr_2  \gx_2  (1-\gx_1)^{-1} \} (\bp_1{}+\bp_2)
 + \{\gr_1 \gx_2(1-\gx_3)^{-1} + \gx_1 \}(\bp_3{}+\bp_2) ]^\pa\,.
 \eee

 Since   $d Z$-dependent terms do not contribute
 to this sector (are weak), $\dr$-exact terms \eq{2DB3bound=} do not contribute to the final result as well.

As  mentioned above,
the terms of   \eq{2ObratnoB3bound} generated by
$
\dr\big\{
     \theta(\gr_1)\theta(\gr_2)\big\}$  cancel against non-zero 'boundary'  terms  of
     \eq{DB3bulkOBRATNOCOCwCCpolu}  through   \eq{ratioksibgt}-like changes of variables.
The rest non-zero terms of \eq{2ObratnoB3bound} generated by
$\dr\big\{
     \theta(\gx_1)\theta(\gx_3)\big\}$ are considered in Section  {\bf E.8}.

Note that the cohomology terms \eq{2COHOMDB3} are presented in Section \ref{fcohomoloCwCCC}
 as those dependent on $\Omega\,, \bar\Omega$ \eq{CgoCCw3B3r=}  and \eq{31w4B3r=}.
\subsection*{E.6 \quad $\dr_x B_2 |_{C\go  CC}$  }
\addcontentsline{toc}{subsection}{E.6 \quad $d_x B_2 |_{C\go  CC}$  }
 \label{dB2CgoCC}
By virtue of \eq{C2gen}, taking into account \eq{UpsgoCC} and  \eq{UpsCgoC} and their conjugates, one obtains from \eq{B2f} and \eq{bB2f}
\bee \label{dxB2+dxbB2CgoCC}
&&\dr_x \big(  B^{\bar\eta}_2 +B^{ \eta}_2\big) |_{C\go  CC}=
\ff{\eta\bar{\eta}}{16}\!\ls \!\int\limits_{\gt \bgt\gx(2) \gs(2)\gr(2) }
\ls    \mu\,
  \Big[ \tr(\bgt )l(\gt)\big\{- \EEE(\wmv_1|\bwmv_1)+
    \EEE(\wmv_2|\bwmv_2)\big\} \\ \nn
&&  +\tr(\gt )l(\bgt)\big\{\EEE(\wmv_3|\bwmv_3)
  - \EEE(\wmv_4 |\bwmv_4 )\big\}\Big]\,C \go CC k\bar{k}\,
\eee
 with $\mu = \nabla(\gs(2))\nabla(\gr(2)) \nabla(\gx(2)) $
  and
   \bee\label{dxB2CgoCC1} &&\wmv^\ga{}_{1} = \tau z^\ga -(1-\gt)
[ \gs_2   p_0 {}  -\gs_1   p _1 {} +\gs_2   p _2 {}  + \gs_2   p_3{} ]^\ga\q \\ \nn
&& \bwmv^\pa{}_1  =
-[- (\gx_1\gr_2  +\gr_1 ) \bp_0 -\bp_1-\gx_1  \bp_2+ \gx_2 \bp_3
     ]^\pa \,,
 \\\label{dxB2CgoCC2}
&&\wmv^\ga{}_{2} = \tau z^\ga -(1-\gt)
[ -\gs_1   p_0 {} -\gs_1   p _1 {} -\gs_1   p _2 {}  + \gs_2   p_3{}  ]^\ga \q\\ \nn
&& \bwmv^\pa{}_2 =
-[  (-\gx_1\gr_1  +\gx_2\gr_2 ) \bp_0   -\gx_1  \bp _1
+\gx_2 \bp_2 +    \bp_3 ]^\pa \,,
\\
\label{dxB2CgoCC3}
&&\wmv^\ga{}_{3} =
-[- (\gx_1\gr_2  +\gr_1 ) p_0- p_1-\gx_1  p _2 + \gx_2 p_3   ]{}^\ga  \q \\ \nn
&& \bwmv^\pa{}_3  =  \bgt \bz^\pa -(1- \bgt)[\gs_2 \bp_0-\gs_1  \bp _1+ \gs_2(\bp_2+\bp_3) ]^\pa
  \,,\\\label{dxbB2CgoCC4}
&&\wmv^\ga{}_{4} =
-[ - (\gx_1\gr_1 -\gx_2\gr_2 ) p_0 -\gx_1p_1+\gx_2 p_2 +p_3 ]{}^\ga\q   \\ \nn
&& \bwmv^\pa{}_4  =  \bgt \bz^\pa -(1- \bgt)[ - \gs_1\bp_0   -\gs_1  \bp _1-\gs_1 \bp_2
+ \gs_2\bp_3 ]^\pa
  \,.
 \eee

These terms are considered in Section  {\bf E.8}.

 \subsection*{E.7 \quad Rest 'bulk' terms}
\addcontentsline{toc}{subsection}{E.7 \quad Rest 'bulk' terms}
\label {restE7}

Taking into account that the  leftover non-zero  'bulk' terms
of Sections    {\bf E.2} -  {\bf E.4}
(namely, those dependent on $\wmv, \bwmv$
\eq{COW2CgoC1}, \eq{wbarW1B2CgoCC}, \eq{12Ugo21w1B3}\,, \eq{COW2CgoC2}\,,
   \eq{wW1barB2=CgoCC}\,, \eq{2Ugo21w1B3}) are spin-local,
   one can straightforwardly make sure that the sum of these terms
    equals to a total  differential that gives zero modulo the
     'boundary' terms \eq{poluCOrestbulk}:
 \bee\label{restbulk}
    &&
   \ff{\eta\bar{\eta}}{16}
\int\limits_{\gt  \bgt  \gr(2)   \gx(2)   \gs(2)  } \ls
 \Big\{  \dr \Big[\Box (\gt,\bgt)  \nabla(\gs(2))\nabla(\gr(2)) \nabla(\gx(2)) \Big]
 \\ \label{poluCOrestbulk}&&-
    \dr \Big[\Box (\gt,\bgt) \Big]  \nabla(\gs(2))\nabla(\gr(2)) \nabla(\gx(2))       \Big\}
        \Big\{\EEE ( \wmv_1 |\bwmv_1 )- \EEE( \wmv_2 |\bwmv_2 )\Big\}
  C\go CC   k\bar{k}\,
,\eee
   \bee \label{UNIFrestleft}
  &&\wmv_1^{ \ga}:= \gt   z{}^\ga- (1-\gt  )[ (\gx_2 \gr_2  \gs_2- 1) p_0 -
  p _1{}+ p_3  - \gs_1 (p_3{}+ p_2)]^\ga  \q
 \\ \nn
&&\bwmv_1^{ \pa}:= \bgt   \bz{}^\pa -(1-\bgt  )[ (\gx_1\gr_1+\gx_2 )\bp_0    +
(\bp_3{}+\bp_2) ]^\pa\,,
\\
\label{UNIFrestrightgt}  && \wmv_2^{ \ga}:= \bgt    z{}^\ga -(1-\gt  )[ (\gx_1\gr_1+\gx_2 )p_0    +
( p_3{}+p_2)^\ga]\q
 \\ \nn
&&\bwmv_2^{ \pa}:= \gt    z^{ \ga} -(1-\gt )[ (\gx_2 \gr_2  \gs_2- 1) \bp_0 -
  \bp_1{}+\bp_3  - \gs_1 (\bp_3{}+ \bp_2)]^\pa \,.
 \eee
 Note, that the terms \eq{restbulk} and \eq{poluCOrestbulk}  are spin-local.

Hence, at this stage, all 'bulk' terms cancel. We are left with the
 'boundary'  terms of \eq{poluCOrestbulk}. The non-zero ones are proportional to
 $\tr(\gt)$ or $\tr(\bgt)$, namely  those dependent on
  \bee \label{UNIFrestleftgt0}
  &&\wmv_1^{ \ga}|_{\gt=0}= -[ (\gx_2 \gr_2  \gs_2- 1) p_0 -
  p _1{}+ p_3  - \gs_1 (p_3{}+ p_2)]^\ga  \q
 \\ \nn
&&\bwmv_1^{ \pa} = \bgt   \bz{}^\pa -(1-\bgt  )[ (1-\gx_1\gr_2 )\bp_0    +
(\bp_3{}+\bp_2)]^\pa\,
\,, \\ \label{UNIFrestrightgt0}
   && \wmv_2^{ \ga}:= \gt    z{}^\ga -(1-\gt  )[ (1-\gx_1\gr_2 )p_0    +
( p_3{}+p_2)]^\ga\q
 \\ \nn
&&\bwmv_2^{ \pa}|_{\bgt=0}= -[ (\gx_2 \gr_2  \gs_2- 1) \bp_0 -
  \bp_1{}+\bp_3  - \gs_1 (\bp_3{}+ \bp_2)]^\pa \,,
 \eee
 are considered in the next Section.
\subsection*{E.8 \quad Rest cohomology terms}
\addcontentsline{toc}{subsection}{E.8 \quad Rest cohomology terms}
\label{LastCgoCCgHOM}
\label{Nu POCH}

 Now we are in a position  to consider non-zero 'boundary'  terms  of Sections
 {\bf E.1}, {\bf E.2}, {\bf E.5}, {\bf E.6} and {\bf E.7}
  contained in Eqs.~\eq{CW2CgoCCbound}, \eq{W2goCCboundgt}, \eq{2ObratnoB3bound},
  \eq{dxB2+dxbB2CgoCC} and \eq{poluCOrestbulk}.

For instance, consider the    terms with $\gt=0$, \ie %$\sim \tr(\gt)$.
 \\
1.
  Eq.~\eq{CW2CgoCCbound} with $\wmv, \bwmv$  \eq{W2goCCw2}.\\
2. Eq.~\eq{W2goCCboundgt} with $\wmv, \bwmv$  \eq{COW2CgoC3}.
\\
3. The  term  of Eq.~\eq{2ObratnoB3bound}, generated by
$\wmv, \bwmv$ \eq{31w4B3rCwCC}   proportional to $\tr(1-\gx_1-\gx_2 )
\tr (\gx_3)$ with
\bee\label{ObratnoB3boundgt01} &&
  \wmv^\ga=-[  (\gs_1(1-\gx_1)-(1 )) p_0  -    (p _1{}+ p_2) + (1-\gx_1)(p_3{}+ p_2) ]^\ga  \q
 \\ \nn
&&\bwmv^\pa = \bgt   \bz{}^\pa -(1-\bgt  )[  (\gs_1\{-\gr_2 \gx_2 +  1 \}
-    \gr_2     ) \bp_0
 -   \gr_2  (\bp_1{}+\bp_2)
 + \{\gr_1 \gx_2  + \gx_1 \}(\bp_3{}+\bp_2) ]^\pa\,.
 \eee
   4. The  non-zero term  of Eq.~\eq{2ObratnoB3bound}, generated by
$\wmv, \bwmv$  \eq{31w4B3rCwCC}    proportional to \\$\tr(1-\gs_1-\gx_2-\gx_3 )\times
\tr (\gx_1)$ with \bee\label{ObratnoB3boundgt02} &&
 \wmv^\ga =-[  (\gs_1 - \gx_2) p_0  -  \gx_2  (p _1{}+ p_2) +  (p_3{}+ p_2) ]^\ga  \q
 \\ \nn
&&\bwmv^\pa = \bgt   \bz{}^\pa -(1-\bgt  )[  (\gs_1 \gr_1
-\{  \gx_3+\gr_2  \gx_2  \} ) \bp_0
%\\ &&\nn
- \{  \gx_3+\gr_2  \gx_2    \} (\bp_1{}+\bp_2)
 +  \gr_1    (\bp_3{}+\bp_2) ]^\pa\,.
 \eee
 5. Eq.~\eq{dxB2+dxbB2CgoCC} with $\wmv, \bwmv$  \eq{dxB2CgoCC3} \,.
\\ 6. Eq.~\eq{dxB2+dxbB2CgoCC} with $\wmv, \bwmv$  \eq{dxbB2CgoCC4}\,.
\\7.   Eq.~\eq{poluCOrestbulk}   with $\wmv, \bwmv$  \eq{UNIFrestleftgt0}\,.
\\

 I. Firstly, we observe that the sum of the terms Eq.~\eq{W2goCCboundgt} with $\wmv, \bwmv$  \eq{COW2CgoC3},
   Eq.~\eq{2ObratnoB3bound} with $\wmv, \bwmv$
\eq{ObratnoB3boundgt02} and
 Eq.~\eq{dxB2+dxbB2CgoCC} with $\wmv, \bwmv$  \eq{dxbB2CgoCC4}
  acquires the form
\bee \label{Coh3UCgoCC}
 &&
 \ff{\eta\bar{\eta}}{16}\!\ls \!\int\limits_{\gt \bgt\gb(2)\ga(2) \gs(2)\gr(2) }
\ls\ls     \!\!
       \Big(\dr[\tr(\gt)\,\mu \,l(\bgt) ]- \tr(\bgt)\tr(\gt)
       \mu \Big)    \EEE( \wmv |\bwmv )\Big] C \go CC k\bar{k},\qquad
\eee
where
$\mu=\nabla(\gb(2))\nabla(\gr(2))\nabla(\gs(2))\nabla(\ga(2))      \q $
  \bee \label{EX03}  &&
  \wmv^\ga = -
( [ \gr_2 -\gs_1  ] p_0-\gs_1 p_1+\gs_2 p_2 + p_3)^\ga
\\ \nn
&&\bwmv^\pa = \bgt  \bz^\pa -(1-\bgt )
\{  \gb_1(  \ga_1[ \gs_1    -\gr_1   ] \bp_0-\ga_1\gs_2\bp_1+\ga_1\gs_1\bp_2 +
 \bp_3)+\\ \nn &&
 \gb_2(-\ga_1\bp_0    -\ga_1  \bp _1-\ga_1 \bp_2
+ \ga_2\bp_3)
  \}^\pa \,.
\eee

Since \eq{EX03} was obtained by application of IH to
\eq{dxbB2CgoCC4} and \eq{COW2CgoC3},  the cohomology term  of \eq{Coh3UCgoCC}
satisfies the   condition \eq{nerav}, \ie is MNL.
It is represented  on the   \rhs of Eq.~\eq{ResultCgoCCb1}  $\wmv_3\,,\bwmv_3$ \eq{Cohgt03=}.
\\

II. Secondly, noticing that
 the following   expression is exact, thus giving zero in the  $d Z$-independent sector
 upon integration,
 \bee \label{dCoh1UCgoCC0}
&&
\ls\ls -\ff{\eta\bar{\eta}}{16}\!\ls \!\int\limits_{\gt \bgt\gb(2)\gs(2) \ga(2)\gr(2) }\ls
\ls     \!\!
      \dr \Big( \tr(\gt)\mu \theta(\bgt)\theta(1-\bgt)     \EEE( \wmv |\bwmv )\Big)\,C \go CC k\bar{k}\,,
\eee
with  $\mu =\nabla(\gb(2))\nabla(\gr(2))\nabla(\gs(2))\nabla(\ga(2))$
and   \bee \label{EX01}
&&\wmv^\ga  =
[- (-\gr_2\gs_2  + 1 ) p_0-p_1-\gr_1  p _2 + \gr_2 p_3   ]{}^\ga    \\ \nn
&& \bwmv^\pa{}  =  \bgt \bz^\pa -(1- \bgt)[
\gb_1\ga_{1 } \bp_0-\gb_1\ga_{2 }  \bp _1+ \gb_1\ga_1(\bp_2+\bp_3)+\\ \nn&&
 \gb_2 (\gs_2-\gs_2 \ga_2 \gr_2
-    \ga_2     ) \bp_0
 - \{ \gb_2 \ga_2   \} (\bp_1{}+\bp_2)
 + \gb_2\{-\ga_2 \gr_2  +  1 \}(\bp_3{}+\bp_2)]^\pa
  \,,\eee
  we observe that the differentiation yields a sum of the
terms of  Eq.~\eq{CW2CgoCCbound} with $\wmv, \bwmv$
  \eq{W2goCCw2},     Eq.~\eq{2ObratnoB3bound}  with $\wmv, \bwmv$\, \eq{ObratnoB3boundgt01},
  Eq.~\eq{dxB2+dxbB2CgoCC}  with $\wmv, \bwmv$ \eq{dxB2CgoCC3}
  plus the following one:
   \bee \label{dCoh1UCgoCC}
&&
  - \ff{\eta\bar{\eta}}{16}\!\ls \!\int\limits_{\gt \bgt\gb(2)\gx(2) \ga(2)\gr(2) }\ls
     \!\!
        \tr(\gt)\nabla(\gb(2))\nabla(\gr(2))\nabla(\gs(2)) l(\bgt)       \EEE( \wmv |\bwmv )\Big)\,C \go CC k\bar{k}\,
\eee
with
   \bee\label{Obratno1gt0}&&\wmv^\ga{}  =
[- (-\gr_2\gs_2  + 1 ) p_0-p_1-\gr_1  p _2 + \gr_2 p_3   ]{}^\ga \,,   \\ \nn
&& \bwmv^\pa{}   =  \bgt \bz^\pa -(1- \bgt)[
  (1- \gb_2  \gs_1    ) \bp_0
  +  (\bp_3{}+\bp_2)]^\pa
  \, \eee
  plus  the   cohomology term represented  with the  minus sign
	in  Eq.~\eq{ResultCgoCCb1}
with   $\wmv \,,\bwmv $ \eq{Cohgt01=}.

 Note that to obtain
\eq{EX01} we  apply IH to
   \eq{dxB2CgoCC3}\,    and
\eq{ObratnoB3boundgt01}, hence preserving MNL.
\\

III. Finally, applying IH
 to the  terms
 \eq{dCoh1UCgoCC}  with $\wmv, \bwmv$    \eq{Obratno1gt0}
   and \eq{poluCOrestbulk} with $\wmv, \bwmv$ \eq{UNIFrestleftgt0}, we obtain   the following
  exact form that does not contribute to the vertex
 \bee \label{Coh2UCgoCCb}
 &&
  \ff{\eta\bar{\eta}}{16}\!\ls \!\int\limits_{\gt \bgt\gb(2)\ga(2) \gs(2)\gr(2) }
 \ls    \!\!
       \dr\Big(\tr(\gt)\mu \theta(\bgt)\theta(1-\bgt)     \EEE( \wmv |\bwmv ) \Big)\,C \go CC k\bar{k}\,
       \eee
with $\mu=\nabla(\gb(2))\nabla(\gr(2))\nabla(\gs(2))\nabla(\ga(2))
 $
 and
   \bee
  \label{EX02b}&&\wmv^\ga:= -[ (\ga_1\gb_2 \gs_2  \gr_2+\ga_2\gr_2\gs_2- 1) p_0 -
  p _1{}+ p_3  - \gr_1 (p_3{}+ p_2)]^\ga  \q
 \\ \nn
&&\bwmv^\pa:= \bgt   \bz{}^\pa -(1-\bgt  )[ (1-\gb_2\gs_1 )\bp_0    +
(\bp_3{}+\bp_2)]^\pa\,
\,. \eee

 The sector of     terms with $\bgt=0$   is considered  analogously. The final results are
 presented on the \rhs of Eq.~\eq{ResultCgoCCb1} with  $\wmv \,,\bwmv $ \eq{Cohgt03b}
and   \eq{Cohgt02b}.

\addcontentsline{toc}{section}{\,\,\,\,\,\,\,References}
\section*{}

  \end{document}